\documentclass[superscriptaddress, aps, pre,twocolumn]{revtex4-2}
\usepackage{amsmath,amssymb}
\usepackage{graphicx,hyperref,color}
\usepackage{bm}

\begin{document}

\title{HTMPC: A heavily templated C++ library for large scale particle-based mesoscale hydrodynamics simulations using multiparticle collision dynamics}

\author{Elmar Westphal}
\affiliation{Peter Gr\"unberg Institute and J\"ulich Centre for Neutron Science, Forschungszentrum J\"ulich, 52425 J\"ulich, Germany}
\author{Segun Goh}
%\email{s.goh@fz-juelich.de}
\affiliation{Theoretical Physics of Living Matter, Institute for Advanced Simulation, Forschungszentrum J{\"u}lich, 52425 J{\"u}lich, Germany}
\author{Roland G. Winkler}
%\email{r.winkler@fz-juelich.de}
\affiliation{Theoretical Physics of Living Matter, Institute for Advanced Simulation, Forschungszentrum J{\"u}lich, 52425 J{\"u}lich, Germany}
\author{Gerhard Gompper}
\email{g.gompper@fz-juelich.de}
\affiliation{Theoretical Physics of Living Matter, Institute for Advanced Simulation, Forschungszentrum J{\"u}lich, 52425 J{\"u}lich, Germany}
\date{\today}
\begin{abstract}
We present HTMPC, a Heavily Templated C++ library for large-scale simulations implementing  multi-particle collision dynamics (MPC), a particle-based mesoscale hydrodynamic simulation method. The implementation is plugin-based, and designed for distributed computing over an arbitrary number of MPI ranks. By abstracting the hardware-dependent parts of the implementation, we provide an identical application-code base for various architectures, currently supporting CPUs and CUDA-capable GPUs. We have examined the code for a system of more than a trillion MPC particles distributed over a few thousand MPI ranks (GPUs), demonstrating the scalability of the implementation and its applicability to large-scale hydrodynamic simulations. As showcases, we examine passive and active suspension of colloids, which confirms the extensibility and versatility of our plugin-based implementation. \\ \\
\emph{Keywords:} Mesoscale hydrodynamic simulation; Multiparticle collision dynamics (MPC)
\end{abstract}

\maketitle

\section{Introduction}

Solid bodies immersed in fluids experience hydrodynamic forces by local fluid motion, which can be of thermal origin on micro- and sub-micrometer scales or initiated by the motion of other immersed objects. Such hydrodynamic interactions are long range in nature and therefore determine the dynamical and structural properties of suspensions. They are particularly relevant for soft matter systems, which encompass traditional complex fluids such as amphiphilic mixtures, colloidal suspensions, polymer solutions, but also biological matter~\cite{gomp:09}. Soft matter systems are characterized by a wide range of relevant length and time scales -- from the atomic scales of solvent particles to the mesoscale of colloidal particles, and the macroscopic scale of phase transitions -- and they are sensitive to thermal fluctuations and to weak perturbations~\cite{dhon:08}. Theoretical and simulation studies reveal the influence of hydrodynamic interaction on the dynamics of polymers~\cite{doi:86,huan:10,thee:16} and colloids~\cite{dhon:96,padd:06} in solution in absence and presence of flow fields~\cite{huan:10,chel:12}. Similarly, biological fluids~\cite{wan:23} share many rheological properties with soft-matter systems. For example, in blood flow, not only the collisions between the deformable red blood cells but also the hydrodynamic interactions due to flow of the embedding blood plasma are essential~\cite{nogu:05,fedo:10.1}. Cellular flows may also involve active components, so that their behavior can then be modeled as active fluids. Nature provides a plethora of autonomous microswimmer~\cite{laug:09,gomp:16}, particular examples are flagellated bacteria~\cite{hu:15,mous:20} and sperm cells~\cite{elge:15}, which propel themselves via hydrodynamic interactions.

%Inherently, soft matter systems are characterized by a wide range of relevant length and time scales, from the atomic scales of solvent particles to the mesoscale of colloidal particles, and the macroscopic scale of phase transitions. Examples include polymer solutions, colloidal suspensions, and liquid crystals. Particularly, hydrodynamic interactions mediated the embedding fluid in such solutions governs the structural and dynamical properties of the dissolved objects~\cite{doi:86,dhon:96,ripo:05,huan:10,thee:16}. Similarly, biological fluids~\cite{wan:23} involve multi-scale structural components, and share many rheological properties with soft-matter systems. For example, in blood flow, not only the collisions between the deformable red blood cells but also the hydrodynamic interactions due to flow of the embedding blood plasma are essential~\cite{nogu:05,fedo:10.1}. Cellular flows may also involve active components, so that their behavior can then be modeled as active fluids. In many cases, suspended objects are capable of active swimming, e.g., in bacterial and sperm suspensions~\cite{elge:15,mous:20}, which relays on hydrodynamic interactions.

The presence of the disparate time, length, and energy scales in soft matter systems renders the application of conventional (all atom) simulation techniques challenging. To mimic the behavior of atomistic systems on the length scale of the mesoscopic objects requires the application of ``coarse-grained'' or mesoscopic approaches~\cite{gomp:09}. Accordingly, mesoscale simulation methods have been developed such as Dissipative Particle Dynamics (DPD)~\cite{hoog:92,espa:17}, Lattice Boltzmann (LB) \cite{mcna:88,succ:01,duen:09}, Direct Simulation Monte Carlo (DSMC) \cite{bird:94,plim:19}, and multiparticle collision dynamics (MPC)~\cite{kapr:08,gomp:09}.

Among the mesoscale simulation techniques, MPC is particularly suitable for simulations on massively parallel GPU-based supercomputers~\cite{west:14,howa:18}. The main advantage is that the interactions and momentum exchange between the fluid particles occur locally in collision cells, which renders the calculation of large inter-particle distances obsolete, but still gives rise to long-range hydrodynamic interactions~\cite{huan:12,thee:16}. In addition, the streaming of every fluid particles is carried out independently. Moreover, MPC incorporates thermal fluctuations by construction. Therefore, MPC is suitable for massive parallel computing and can be utilized to investigate large thermal systems beyond scales that a single processor can handle.

Since the original formulation in 1999~\cite{male:99}, several variants and revisions of MPC have been proposed~\cite{gomp:09}, and it has been applied to a wide range of physical problems. Particularly,  the transport properties of MPC fluids have been obtained theoretically~\cite{ihle:03,ihle:03.1,kiku:03,nogu:08,wink:09} and its hydrodynamic correlation function has been derived~\cite{huan:12}. The applications cover such diverse aspects as sedimenting colloidal suspensions~\cite{padd:04,hech:06} and polyelectrolyte electrophoresis~\cite{fran:08}, liquid crystals~\cite{mand:19}, polymeric systems and protein solutions~\cite{male:00,muss:05,ali:06,niko:10,huan:10,huan:13,chen:17,niko:17,lieb:20,deva:22,zwan:23}, ferro- \cite{ilg:22} and nanofluids \cite{wang:23}, and even self-propelled particles~\cite{thak:12,thee:16.1,zant:22,qi:22,goh:23}, active nematics~\cite{dura:23} and active polymers~\cite{jain:22,clop:22}, as well as biological systems such as suspensions of bacteria~\cite{hu:15,hu:15.1,eise:16.1,mous:20,ning:23}, sperm~\cite{elge:10,chin:18,rode:19}, and parasites, e.g., trypanosomes~\cite{hedd:12} and \emph{Plasmodium falciparum}~\cite{lans:18}.

In this article, we present a highly efficient parallel implementation of MPC, which is suitable for both CPU- and GPU-based computers, and has been evolved from the previous GPU-based implementation presented in Ref.~\cite{west:14}. Our new code has been tested up to $1.5$ trillion MPC fluid particles on $3,540$ NVIDIA A100 GPUs. %We have also made the code available on ... 

\section{Multiparticle collision dynamics}

%\subsection{Fluid}
In MPC, a fluid is represented by $N_s$ point particles of mass $m$ with the positions ${\bm r}_i$ and the velocities ${\bm v}_i$ ($i=1,\ldots, N_s$). Their time evolution is described by a two-step dynamics of alternating  streaming and collision steps. In the streaming step, the MPC particles move ballistically according to 
\begin{align} \label{eq:str}
{\bm r}_i (t+h) = {\bm r}_i(t) + h{\bm v}_i,
\end{align}
where $h$ is the MPC time step. The inter-particle interactions are modeled via a momentum conserving stochastic process in the collision step~\cite{male:99,kapr:08,gomp:09}. Here, all MPC particles are sorted into the cells with side length $a$ of a simple cubic lattice, each of which contains on average $\langle N_c \rangle$ particles. Then, in the stochastic rotation dynamics (SRD) variant of MPC~\cite{gomp:09}, the relative velocities ${\bm v}_{i,{\rm c}} \equiv {\bm v}_i - {\bm v}_{\rm cm}$ of the particles within a cell with respect to the center-of-mass velocity  ${\bm v}_{\rm cm} \equiv (\sum_{i=1}^{N_c} {\bm v}_i)/N_c$ of the cell are rotated by a fixed angle $\alpha$ around a randomly oriented axis according to
\begin{align} \label{eq:col}
{\bm v}_i (t+h) = {\bm v}_{\rm cm}(t) + \mathbf{D}(\alpha) {\bm v}_{i,{\rm c}}(t),
\end{align}
where $\mathbf{D}$ denotes the rotation matrix. The orientation of the rotation axis is independently chosen for every cell and collision step. Notably, the entire collision lattice is randomly shifted before every collision step to insure Galilean invariance~\cite{ihle:01,ihle:03}. 

Several theoretical and technical advances have been achieved in the course of the development of MPC. Here we summarize some of the important issues, which have also been taken into account in our implementation.

\paragraph*{\bf Angular-momentum conservation.} 
Violation of angular-momentum conservation may lead to non-physical torques due to non-symmetric components in the stress tensor~\cite{goet:07,thee:15,yang:15}. To avoid such artifacts, cell-level angular-momentum conservation can be enforced (MPC-SRD+a) by adding an additional constraint to the velocity in Eq.~\eqref{eq:col} at each collision step \cite{nogu:07,thee:15},
\begin{align} \label{eq:col+a}
{\bm v}_i (t+h) = {\bm v}_{\rm cm} + \mathbf{D}(\alpha){\bm v}_{i,{\rm c}} + {\bm \omega}_{\rm c} (t) \times {\bm r}_{i, {\rm c}},
\end{align}
where ${\bm r}_{i,{\rm c}} \equiv {\bm r}_i - {\bm r}_{\rm cm}$ is the relative position with respect to the center-of-mass position ${\bm r}_{\rm cm}$, and the angular velocity is given as
\begin{align} \label{eq:col_ang}
{\bm \omega}_{\rm c} = m \mathbf{I}_{\rm c}^{-1} \sum_{i=1}^{N_c} \left( 
{\bm r}_{i,{\rm c}} \times ({\bm v}_{i, {\rm c}} -\mathbf{D}(\alpha) {\bm v}_{i, {\rm c}})\right).
\end{align}
Here, $\mathbf{I}_{\rm c}$ is the moment-of-inertia tensor of the particles in the collision cell with respect to the center of mass.

\paragraph*{\bf Thermostat.} 
As the MPC-SRD algorithm given by Eq.~\eqref{eq:col} conserves energy, the fluid shows thermodynamic behavior of a microcanonical ensemble. Whenever a canonical ensemble is appropriate, i.e., a constant temperature is desired, the system has to be thermostatted. This is achieved by a Maxwell-Boltzmann scaling (MBS) approach~\cite{huan:10.1}, which rescales the relative velocities in the cells after each collision step via the scaling factor $\kappa$ 
\begin{align}
{\bm v}'_i = {\bm v}_{\rm cm} + \kappa {\bm v}_{i,{\rm c}}.
\end{align}
The scaling factor itself is determined from the kinetic energy of the particles within a cell, i.e.,
\begin{align}
\kappa = \sqrt{\frac{2E_k}{\sum_i m{\bm v}_{i,{\rm c}}^2}},
\end{align}
where the random kinetic energy $E_k$ suffices the distribution function 
\begin{align}
P(E_k) = \frac{1}{E_k \Gamma (f/2)} \left( \frac{E_k}{k_{\rm B}T} \right)^{f/2} \exp{\left( -\frac{E_k}{k_{\rm B}T} \right)},
\end{align}
with $f=3(N_c -1)$ the number of degrees of freedom and $\Gamma$ the gamma function.

\paragraph*{\bf Boundary conditions.} 
In addition to periodic boundary conditions, the fluid can be in contact with confining walls and surfaces of embedded particles such as colloids. Slip as well as no-slip boundary conditions can be employed in MPC. During the streaming step, collisions between MPC particles and boundaries result in a momentum change. Specifically, all the MPC particles which have crossed boundaries are translated back onto the boundaries. For no-slip boundary conditions, the bounce-back rule is applied, where the velocity is simply reverted, i.e., ${\bm v}_i \to -{\bm v}_i$.  For slip-boundary conditions, specular reflection is employed. In case of, e.g., interactions with colloidal particles, momentum conservation implies a change of the linear momentum of MPC particles as well as the linear and angular momenta of the colloid. In the collision step, collision cells are added either in walls or larger immersed particles. The part of these collision cells in walls or immersed particles are filled with uniformly distributed phantom particles, which are identical to MPC particles, such that the whole collision cell contains the same average number $\langle N_c \rangle$ of fluid particles~\cite{lamu:01,wink:09}. This enhances wall-fluid and colloid-fluid interactions and reduces boundary slip~\cite{lamu:01,wink:09}. Specifically, it is required due to the random shift of the collision lattice, otherwise the particle number varies in cells cut by boundaries. The velocities of the phantom particles are taken from a Gaussian distribution with zero mean and variance $\sqrt{k_{\rm B}T/m}$.
%\begin{align}
%{\bm v}_i^p =  {\bm v}_i^{\rm ran},
%\end{align}
%where the components of ${\bm v}_i^{\rm ran}$ are random Gaussian variables with zero mean and variance $\sqrt{k_{\rm B}T/m}$. 
These phantom particles participate in the collision step as all other fluid particles.

\paragraph*{\bf External fields.} A constant gravitational force along a coordinate axis in combination with two parallel confining walls leads to a parabolic Poiseuille flow~\cite{cann:08}. Importantly, in this case, the MPC fluid has to be thermalized~\cite{huan:15}.   

\paragraph*{\bf Parameters.} The simulation of fluid properties requires a suitable choice of the MPC parameters~\cite{padd:04,thee:16}. By the nature of the MPC collision process, hydrodynamics in the fluid appears on length scales $\lambda_c$ larger than the collision cell only. The actual length scale itself depends on the collision time step and increases with increasing collision step~\cite{huan:12}. Moreover, in fluids, momentum transport dominates over mass transport. This is expressed by the Schmidt number $S_c=\nu/D$, where $\nu$ is the kinematic viscosity and $D$ the diffusion coefficient of the MPC particles~\cite{ripo:05}. In MPC, a large Schmidt number is achieved for small collision step sizes and large collision angles. Hence, we recommend to choose these values in the range $h \leq 0.1 \sqrt{m a^2/(k_{\rm B}T)}$  and $\alpha > 90^{\circ}$~\cite{huan:12,ripo:05}.   

Proper hydrodynamics for objects embedded in MPC such as colloids or polymers requires an appropriate choice of their size. Because of the condition $\lambda_c/a >1$, a colloid radius should be $R_c/a\geq 3$ (see Sec.~\ref{sec:showcase})~\cite{thee:16}. Similarly, in a simulation of a polymer its bond-length and monomer diameter should be larger than the collision cell size~\cite{muss:05}. 

Due to the point-particle nature, a MPC fluid is compressible with an ideal gas equation of state. The compressibility can be controlled by the average number of fluid particles in a collision cell. For systems with passive colloids or polymers, $\langle N_c \rangle =10$ is recommended. However,  in systems with self-propelled particles, e.g., squirmers, a larger average number $\langle N_c \rangle$ may be necessary to avoid depletion effects, particularly at large concentrations; here, we suggest $\langle N_c \rangle = 50$ for P\'eclet numbers in the range $\rm Pe \sim  10^1 - 10^2$~\cite{thee:18,qi:22,goh:23} (see Sec.~\ref{sec:showcase}).

\section{Implementation}

\subsection{Code Structure}
\paragraph*{\bf Basic structure.}
The code is a C++17 template library. It consists of a class template representing a simulation environment using plugins that contain the simulation data and/or the code to manipulate it. This environment is generated using rules and definitions provided by a descriptor class that is passed as a template argument. This class contains a sequence of plugin templates, an enumeration of partial simulation steps to be executed in every iteration, an arbitrary number of type-declarations and constexpr (compile time) variable-declarations and a sequence of parsers to interpret these declarations. 

\begin{figure*}
	\centering
  	\includegraphics[width= 1.6\columnwidth]{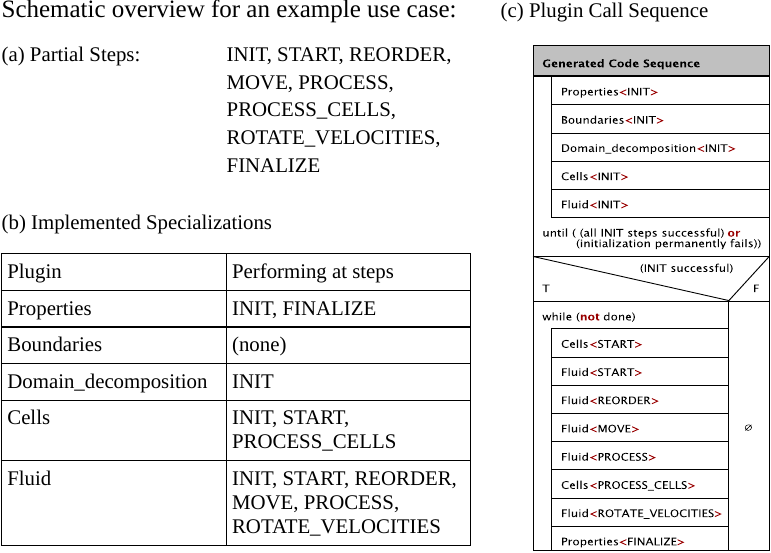}
	\caption{\label{fig:structure}
	Schematic overview for an example use case.
        (a) Partial steps are defined as an enumeration in the configuration class. 
        (b) Plugins containing specialized functor templates to be called at partial steps listed in (a).
        (c) The plugin call sequence ultimately generated at compile time.
}
\end{figure*}

\paragraph*{\bf Partial Simulation Steps.}
In each MPC step a number of operations is performed on the simulation data. To coordinate these operations, each iteration step can be divided into an arbitrary number of partial steps, such as the movement of fluid particles, the generation of rotation matrices, etc. As the partial steps have to be performed in a given order, they are defined as an enumeration. The code allows the user to extend this basic enumeration by inserting an arbitrary number of additional steps into it, provided all existing steps remain in their original order.

\paragraph*{\bf Plugins.}
The data and code for the simulation are stored in plugin class templates. These are listed in a sequence given in the configuration class and are instantiated according to rules also given there. In the generated simulation class, instantiations of all plugin templates specialized by the configuration class are inherited and can be accessed in different, configurable ways. The plugins may contain functor templates specialized for each partial iteration step, containing the code to be performed at that particular step. Besides data and code to be performed regularly at certain times in the iteration step, plugins can also represent boundary conditions by adding additional functor templates that can be applied to particle data in certain situations such as particle movement or processing. The included plugins implement the MPC fluid, MPC collision cells, different boundary conditions, etc. Additional plugins can be added following the given interface.

\paragraph*{\bf Generating iteration steps.}
As described, each iteration step is divided into a number of partial steps. During each iteration step the partial steps will be performed in the order given in their enumeration. During each partial step, all plugins will be traversed in the order given in the plugin sequence and if they contain a functor template specialized for that particular step, the functor will be instantiated to be called. The only exception to this is the initialization step, the functor of which is only called once after the simulation environment is constructed. Only during initialization, it is possible to change the order of execution of the plugins, making it possible to process dependencies between them. Otherwise, all dependencies must be implicitly resolved by the given order of plugins. Regular operation will commence as soon as all plugins are initialized. If the initialization of one or more plugins fails permanently, the program will be aborted.

\subsection{Distributed Simulation}
The code is designed to optionally distribute the workload over an arbitrary number of MPI ranks, which is achieved by splitting the simulation system into spatial subdomains. In addition, particles are reassigned to the subdomains covering their current position at regular intervals. Between these reordering operations, particles that leave their subdomains are temporarily sent to the appropriate subdomain for processing and their updated velocities are imported back.

\subsubsection{Domain Decomposition}
The amount of necessary communication depends on the number of particles exchanged, and therefore, the surface area of the subdomains. For optimal performance, the domain decomposition algorithm minimizes the surface area of subdomains, by generating subdomains close to cubic shapes. To this end, the total number of subdomains given by the number of MPI ranks is factorized at first. The currently longest side of the system will then be subsequently divided by the largest remaining factor in a loop over all factors from the factorization. If a shear flow is considered, the system is not split along the corresponding discontinuous axis to avoid difficulties during particle exchanges resulting from the fact that we can not define a fixed neighborhood relation between subdomains due to the continuous shifts in shear-offset across the boundaries. Lastly, the number of remaining collision cells in each direction, given as the remainder of the system size divided by the number of domains, will be distributed evenly over the subdomain sizes, i.e. a size of 20 splits into 3 will result in partial sizes of (7,7,6).

\subsubsection{Temporary Exchange of Particles for Processing}
\label{sec:temp}
During every iteration step, the respective collision cell of each MPC particle is calculated from the position of particles, the random shift of the current iteration, and applicable boundary conditions. If the collision cell of a MPC particle is outside the subdomain the particle is currently assigned to, then the particle must temporarily be transferred to the respective subdomain and processed there. Each subdomain has one export buffer - similar to the fluid particle storage - for preparing sets of particles to be sent to other domains, and one import buffer for receiving particles from other domains. These buffers only contain the data members necessary for the currently activated feature set.

Whenever a particle needs to be exported, the counter of particles to be sent from the current subdomain to the destination subdomain is incremented, and a data triplet of the particle index, destination subdomain, and previous export count (as a destination-relative index) is appended to a vector. After finding all particles to be exported, the prefix sum of the export counts of destination subdomains is calculated, which represents the offsets for each destination subdomain in the output buffer. Then the vector of data triplets is traversed and the data of each particle contained in the traversed vector is copied to the export buffer position calculated by adding the destination offset from the prefix sum and the destination-relative index of the particle. In this way, a buffer containing particle data is ordered by their respective destination subdomains, which is efficient especially on GPUs, as each step can be performed in parallel and the export buffer for all domains can be built with a single kernel call.

In almost all cases fluid particles are received from direct neighbors and the exchange is done using a \texttt{MPI\_Neighbor\_alltoall} collective call. Otherwise, a more expensive fallback path using \texttt{MPI\_Alltoall} is used. Then a prefix sum is calculated over the import counts, resulting in the import offsets for each subdomain. Such obtained import- and export-counts and -offsets are then used in neighbor-collective or discrete MPI-calls to exchange the necessary data members, which can be done without explicitly transferring data to/from the GPUs by using a CUDA-aware MPI implementation. Now the code is ready to perform all remaining MPC steps, after which the processed particle velocities are transferred back to their original subdomains and copied back from the exchange buffer to their respective particles. 

\subsubsection{Permanent Exchange of Particles During Reordering}
During the course of the simulation, a growing number of particles move away from the subdomain they were originally assigned to, resulting in a higher communication load as well as a growth in buffer size. To avoid such penalty in performance, the code reassigns (or reorders) particles to the domains where they are most likely processed. Reasonable reorder intervals must balance the time spent for reordering particles and the time gained through faster operation, which also depends on the time step length and system architecture. In most setups, an optimal interval lies between ten and a few hundred iteration steps. The algorithm for transferring the particles is the same as described in Sec.~\ref{sec:temp} for temporary particle transfers. After transfer, however, the relocated particles are inserted and deleted to/from the permanent particle sets of the domains. 

\begin{figure}
	\centering
  	\includegraphics[width= 1\columnwidth]{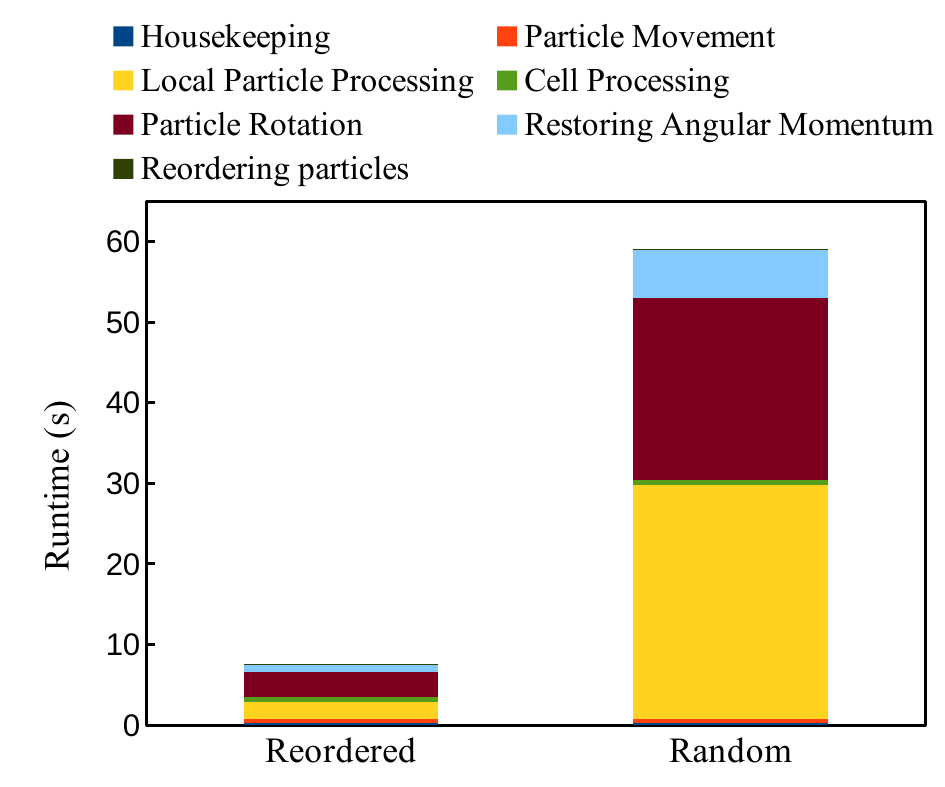}
	\caption{\label{fig:reorder}
	Influence of particle reordering on runtime for 200 iteration steps with AMC. Here, $L=128a$ and $\langle N_c \rangle =10$. The test runs were performed on a GTX2070 GPU.
}
\end{figure}

During the exchange step, the particles in each subdomain are also spatially reordered, which is beneficial for performance in general, as shown in Fig.~\ref{fig:reorder}, and therefore, is also applied to single-rank systems. Reordering the particles in each subdomain proceeds similarly to the filling of export buffers described above, but particles are ordered by collision cells instead of destination subdomains. Particle counts per cell are determined, using the particles still residing in their current subdomain and, in case of distributed systems, particles received from other subdomains in the import buffer. The current particle counts per cell are again used as cell-relative indices and the prefix sum of the final particle counts is used as cell offsets for the reordered particles.

To save memory, reordering is done property-wise, which necessitates only a temporary reorder buffer with the size of the largest data member of the particles (usually the velocity). Specifically, each data member of the currently resident particles is copied to the temporary buffer. The original data storage is resized to fit the data of all remaining resident plus imported particles. Then the data of particles in the temporary buffer that are not exported and those from the newly imported ones in the import buffer are copied to their new positions in the resized main data location. It is also possible to perform this step component-wise for each particle property, to reduce the reorder buffer size by a factor of three, which in turn, corresponds to roughly 15-25\% of the total memory footprint. Due to inefficient memory access patterns and repeated access, however, it comes with a cost of increasing the runtime by a few percent, the exact value of which depends on the reorder intervals.

\subsection{Reproducibility}
Additions are not always associative for floating point values, i.e., $(a+b)+c$ is not necessarily equal to $a+(b+c)$~\cite{gold:91,vill:09}. In a multi-threaded or distributed code, such non-associativity causes diverging results between runs with identical initial conditions, because the order of operations is undefined and also depends on the distribution of data. Even small fluctuations will result in significantly different particle trajectories after a sufficient number of iterations. While each of these runs is still valid, perfectly matching trajectories may be desirable, for justification or debug purposes. 

To achieve this, the code optionally provides bit-perfect reproducibility of particle trajectories by replacing the floating point summations with fixed point summations, where fixed point numbers assign a fixed count of bits to the integral part~\cite{gran:13}. If the resulting range is not defined appropriately, overflows may occur leading to undefined behavior (see e.g., Ref.~\cite{diet:15}) and invalidating the results, which implies an inherent tradeoff between range and precision.
For typical simulations the range of values of the cell parameters is limited and can be well estimated, allowing us to determine the necessary precision for fixed point numbers. The generation of initial particles is also adapted accordingly to achieve perfect reproducibility not only on a single GPU, but also on distributed runs, independent of rank counts or domain decompositions, as long as the same system architecture, compiler, and compiler settings are used. Even though the summations themselves should not be influenced by associated parameters, different hardware components, numerical algorithms or optimizations may introduce subtle changes in the results of other operations. Additionally, the order of particles in memory or in files saved to disk is undefined due to their dependence on the domain decomposition layout and the undefined order of atomic operations.
%\EW{(no, this is a general remark about the limitations of the implementation. We can guarantee that there is a particle with an exactly matching trajectory or, if we provide an particle ID, that the particle with a particular ID will always follow the the same trajectory, but we can not tell at what position in a save-file that particle will be stored, because this depends on the domain decomposition layout and possibly on the undefined order of atomic operations) )}

%\SG{(Maybe not necessary...)} For very small systems, the use of fixed precision additions may slightly limit the achievable stability of the residual momentum of the system, in larger systems the precision is within the same order of magnitude. 

\paragraph*{\bf Fluid Particle ID.}
In the given implementation, unless particle reordering is switched off in an undistributed system, it is not possible to follow certain particles due to the regular exchange of particles between subdomains and/or particle reordering for performance purposes. The code provides an option that assigns an integer tag to each fluid particle, which can be adopted for, e.g., identifying certain particles or color-coding groups. If this option is used, the tag is, by default, generated as a consecutive, unique number for each particle. We note, however, that the assigned ID can also be changed to non-unique numbers because the subdomain of particles and offset within will change unpredictably at regular intervals. Therefore, only in combination with the reproducibility option, a specific ID would apply to a particle with identical properties. 

\subsection{Additional Code Features}

Here additional code features are summarized. More detailed information can be found in Supplementary Information (SI).

\begin{figure}
	\centering
  	\includegraphics[width= 1\columnwidth]{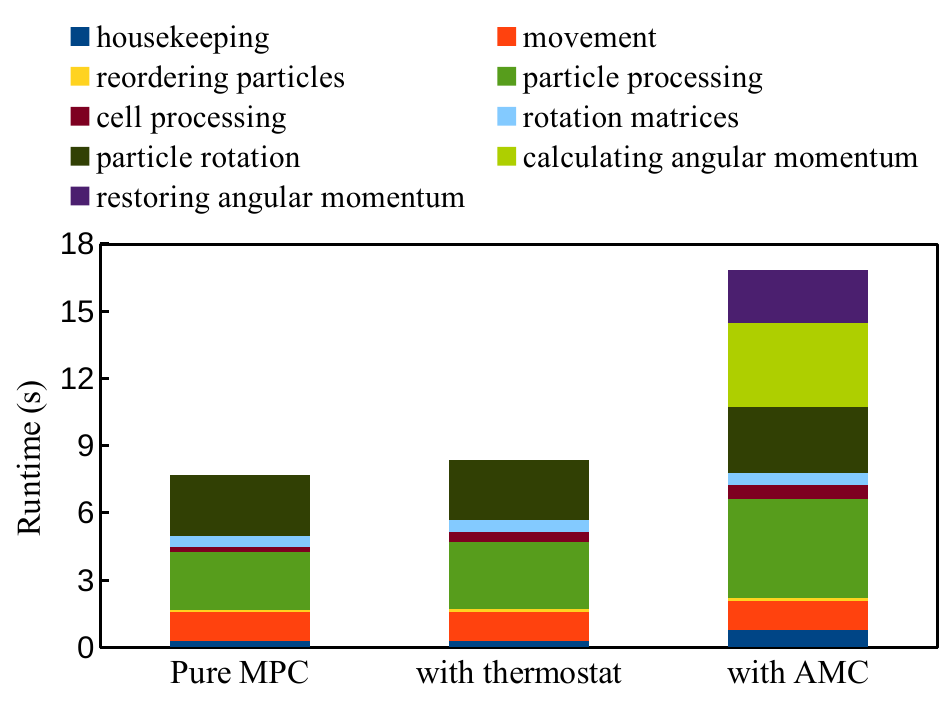}
	\caption{\label{fig:feature}
	Feature dependence of the time spent in partial steps. Here, 1000 iteration steps with $L=100a$ and $\langle N_c \rangle =10$ were performed on a GTX2070 GPU.
}
\end{figure}

\paragraph*{\bf Thermostat.}
The implementation of the Maxwell-Boltzmann scaling (MBS) thermostat \cite{huan:10.1} mainly consists of two components. First, the kinetic energies of all particles are summed up cell-wise during particle processing. Then a scaling factor for each cell is calculated during cell processing, which is applied  to all particle velocities in the cell during the particle rotation. Applying the MBS thermostat increases total calculation time by approximately 10\%, as shown in Fig.~\ref{fig:feature}.

\paragraph*{\bf Local Conservation of Angular Momentum.}
Angular-momentum conservation (AMC) adds new calculation steps and a significant amount of additional calculations to several other steps. Consequently, conserving local angular momentum in combination with the thermostat option will increase calculation times by a factor larger than 2, see Fig.~\ref{fig:feature}. Due to the significantly higher number of operations and their respective error propagation, the residual momentum is roughly two orders of magnitude higher than for plain MPC.

\paragraph*{\bf Boundary Conditions.}
Boundary conditions are implemented as special plugins with additional functionality. Like all other plugins, they may have their own state, data, and code. Additionally, they have a functor template that can be specialized for certain situations, usually when a particle is moved or when it is processed after random shift of the collision-cell lattice was applied. Different boundary conditions can be set for different axes of the simulation system. Using the appropriate interface, boundary condition plugins can be added ad libitum. The included plugins support periodic, slip, and no-slip boundary conditions. Optionally, shear flow can also be added. 
%We note that a plugin implementing cylindrical boundary conditions is also included, though experimental for now.

\paragraph*{\bf Phantom Particles.}
The code provides specialized particle sets that can be used by plugins for the efficient processing of phantom particles. In contrast to standard fluid or solute particles, phantom particles are not subject to movement or distribution, but are created in the subdomain where they are to be processed while random shifts are taken into account. Plugins generating phantom particles initiate their processing, rotation of velocities, and restoration of their local angular momentum at appropriate times, if applicable.

\paragraph*{\bf Gravity.}
It is possible to apply a constant gravitational force parallel to one of the axes. The application can be combined with other options and the implemented boundary conditions.

%\paragraph*{\bf Suspended particles.}
%Blahblah. \SG{my suggestion: keywords could be MD, boundary conditions, slip velocity, and also the shape of particles.} \EW{This section deals with the MPC-related plugins that I would consider the basic building blocks. Suspended particles, especially our squirmer code, are a use-case and information about that should be presented in section V. There is a building block that can be used for point particles in e.g. polymer simulations, that could warrant a short paragraph.)}

\subsection{Hybrid CPU/GPU Implementation}

The code is designed for distributing the system across an arbitrary number of domains, each of which is processed by one MPI rank. The implementation of intra-domain operations favors GPU acceleration, but also works on one CPU core per MPI rank. CPU-based shared memory parallelization like OpenMP for intra-domain operations was also tested, but the performance turned out to be lower than that of a more fine-grained domain-decomposition distribution.

\subsubsection{Identical code base for different architectures}
Most of the code processes either particle- or cell-based operations. Each operation is implemented as a functor which is called once for every cell or particle, and is coded to be hardware agnostic for better maintainability and compatibility between platforms. Hardware dependent operations, possibly leading to atomic operations, are hidden in wrapper classes that are implemented for their designated target architecture. Currently, plain CPU code and CUDA-based GPU acceleration on NVIDIA GPUs are implemented, but the concept is modular and could be extended to other architectures without the need to rewrite any of the actual application code.

\subsubsection{GPU Memory Management}
Accessing CPU memory from a GPU or vice versa is usually not possible, or comes with a significant performance penalty for most of the relevant past and current GPU architectures. In the code, we employ the CUDA-based implementation providing an allocator for so-called Managed Memory, which is a mechanism providing automated data migration on older GPUs or memory paging on more recent hardware, avoiding the need to explicitly migrate data. Still, alternating access to the same data elements from the CPU and GPU leads to implicit data movement and can significantly hurt performance. For instance, access to seemingly independent data residing on the same memory page may cause performance issues related to concurrent access of this page by the application running on the GPU and the MPI library running on the CPU. These issues have been addressed by padding managed memory allocations.

\subsubsection{GPU acceleration}
Code generation and architecture selection is based on C++17 template mechanisms. The only external dependency is an appropriate CUDA compiler. The architecture dependent functionalities that had to be implemented include, among others, an STL-compatible, GPU-based vector template, reduction routines, prefix sums and accumulator classes wrapping atomic additions. As more and more standard library routines are provided in GPU-parallelized versions, some implementations may be updated accordingly in later releases. 

\subsubsection{Atomic Operations}
While early releases of our GPU-based MPC routines used a list-based approach for processing the particles belonging to each collision cell~\cite{west:14}, the current version uses atomic operations for all summations of this kind. Atomic operations used to be very expensive in early GPU architectures, especially for double precision additions which had to be emulated by loops instead. Recent architectures, however, provide native implementations for double precision additions and much faster support for atomic operations in general. Additionally, even with most spatially ordered particles, the list-based access to particle data results in large amounts of data that have to be read multiple times due to the need of reading data blocks of certain sizes (cache lines), while an atomic-operation based implementation allows for coalesced reading patterns. On the other hand, a list-based approach may show better writing patterns if particles are distributed evenly among cells. But overall the atomic-operation based implementation is faster on modern GPUs for our code. Therefore, the list approach was abandoned after ``Kepler'' became the prevalent GPU architecture. 

Certain operations such as binning data for the calculation of optional shear profiles, may result in extraordinary high numbers of concurrent atomic operations on a small amount of data, impairing the performance. In this case, optimized local summation patterns may significantly reduce the number of atomic operations~\cite{west:15}. The code may activate such optimizations, depending on their expected efficiency for certain access patterns and GPU architectures.

\subsubsection{Random Numbers}
For an efficient GPU implementation, we choose the Xorshift algorithm~\cite{mars:03} as the default random number generator (RNG), which can be implemented in a branch- and loop-free manner, while providing good randomness. Specifically, a system is initialized with a main random seed, which can be either passed as a parameter or generated from the random source of the operating system. Using the main random seed only might lead to the same or at least dependent random values for every cell or necessitate an atomic- or lock-based mechanism for accessing the random seed causing a significant performance penalty. Therefore, the cell random seeds are necessary for processing certain parts in parallel, e.g. generating the random rotation matrix per cell. We note that the modular concept of the code also allows users to replace it by a user-provided RNG with a matching interface.
%While the main random seed can be accessed from the plugins, it is important to keep it in sync across MPI ranks. Otherwise catastrophic side effects like different domains using different random shifts will occur.

\subsection{Speed Boundaries}
The computation speed of operations is limited by certain boundaries like the maximum amount of data transferred from/to memory per second or the maximum number of floating point or atomic operations per second that can be performed by the CPU or GPU. Taking into account all particle- and cell-related read/write operations, i.e. fluid particle propagation, at least on GPU, is bound by the memory bandwidth. The combination of the rotation of particle velocities and the calculation of several cell parameters needed for local angular-momentum conservation as implemented in our code is the most expensive step, both calculation- and memory-transfer-wise. Even though it involves most of the algorithm’s atomic operations, it still runs at approximately 60\% of the theoretical memory bandwidth. Without angular-momentum conservation and the atomic operations involved, the rotation of particles is also memory bound.

\subsection{Simulation Data}

\subsubsection{Data Types and Memory Management}
The storage and calculation precision for most values and operations can be adjusted in the configuration class. The default settings are single-precision floating-point numbers for the fluid particle positions and double precision for their velocities~\cite{west:14}. All calculations on cell properties are performed and stored in double precision. Precisions for suspended particles, e.g., in Molecular Dynamics (MD) plugins can be set as necessary. The amounts of data stored for fluid particles and collision cells in the default settings are summarized in Tables~\ref{tab:particle} and~\ref{tab:cell}, respectively. 
%Depending on the architecture, especially for GPU types with precision related performance limitations, changing some operations to lower precision can gain a significant speedup, but it is advised to perform extensive tests to ensure sufficient numerical stability of the results.

%The default precision settings as outlined above lead to the following amounts of data stored for fluid particles and collision cells: TABLE!

\begin{table}
    \centering
\begin{tabular}{ c | c }
\hline
\hline
 $\hspace{0.7cm}$Particle property$\hspace{0.7cm}$ & $\hspace{0.7cm}$Size (bytes)$\hspace{0.7cm}$ \\ 
\hline
\hline
 Position & 12 \\  
 Velocity & 24 \\
 Cell index$^1$ & 4-8 \\
 Reorder buffer$^2$ & 8-24 \\
 ID-tag$^{*,3}$ & 4-8\\
\hline
 Total & 48-76\\
\hline
\hline
\end{tabular}
\caption{\label{tab:particle} The amounts of data for each MPC particle. $^*$ optional; $^1$ mandatory for distributed systems, recommended otherwise, may trade memory footprint for performance; $^2$ depending on maximum system size; $^3$ depending on maximum particle count.}
\end{table}

\begin{table}
    \centering
\begin{tabular}{ c | c }
\hline
\hline
 $\hspace{1.5cm}$Cell property$\hspace{1.5cm}$ & $\hspace{0.7cm}$Size (bytes)$\hspace{0.7cm}$ \\ 
\hline
\hline
 Center of mass velocity & 24 \\  
 Combined mass of particles & 8 \\
 Particle count & 4 \\
 Random seed  & 8 \\
 Energy/thermostat$^{*,1}$ & 8 \\
 Angular momentum$^{*,2}$ & 24 \\
 Center of mass$^{*,2}$ & 24 \\
 Inertia Tensor$^{*,2}$ & 48 \\
\hline
 Total & 44-148 \\
\hline
\hline
\end{tabular}
\caption{\label{tab:cell} The amounts of data for each MPC cell. $^*$ optional; $^1$ if using thermostat; $^2$ if conserving angular momentum.}
\end{table}

%Depending on the feature set and density, a GPU equipped with 40GB of memory can store cubic systems up to $\sim400a$ in size for rho=10 with a minimal feature set and ~200a for rho=50 using the full feature set.
For distributed systems, each subdomain reserves a slightly larger fluid particle capacity to leave room for variations in density between domains. Additionally, import- and export-buffers are needed for particle exchange, whose size depends mostly on the surface area of the domains. As a general rule, the initial buffer size is $\langle N_c \rangle$ multiplied by the number of surface cells, about half of which is caused by the random shift of up to a collision-cell size $a$ for all axes and the remainder to process particles moving out of their assigned subdomain. The number of the latter is reduced by spatially reordering fluid particles at certain intervals. Because memory allocation is an expensive operation, especially on GPUs, the import and export buffers are reserved slightly above the expected size and may be extended by reallocation if needed, but never shrunk. Additional memory may be used by other plugins. One example is boundary conditions representing no-slip surfaces, which need one cell layer to be partially (depending on the random shift) filled with phantom particles for fluid-surface coupling. If the top and bottom no-slip boundaries are located in separate domains, both domains will need this one-cell-layer-wide buffer, otherwise one buffer will be shared.

\begin{figure*}
	\centering
        \includegraphics[width= 2\columnwidth]{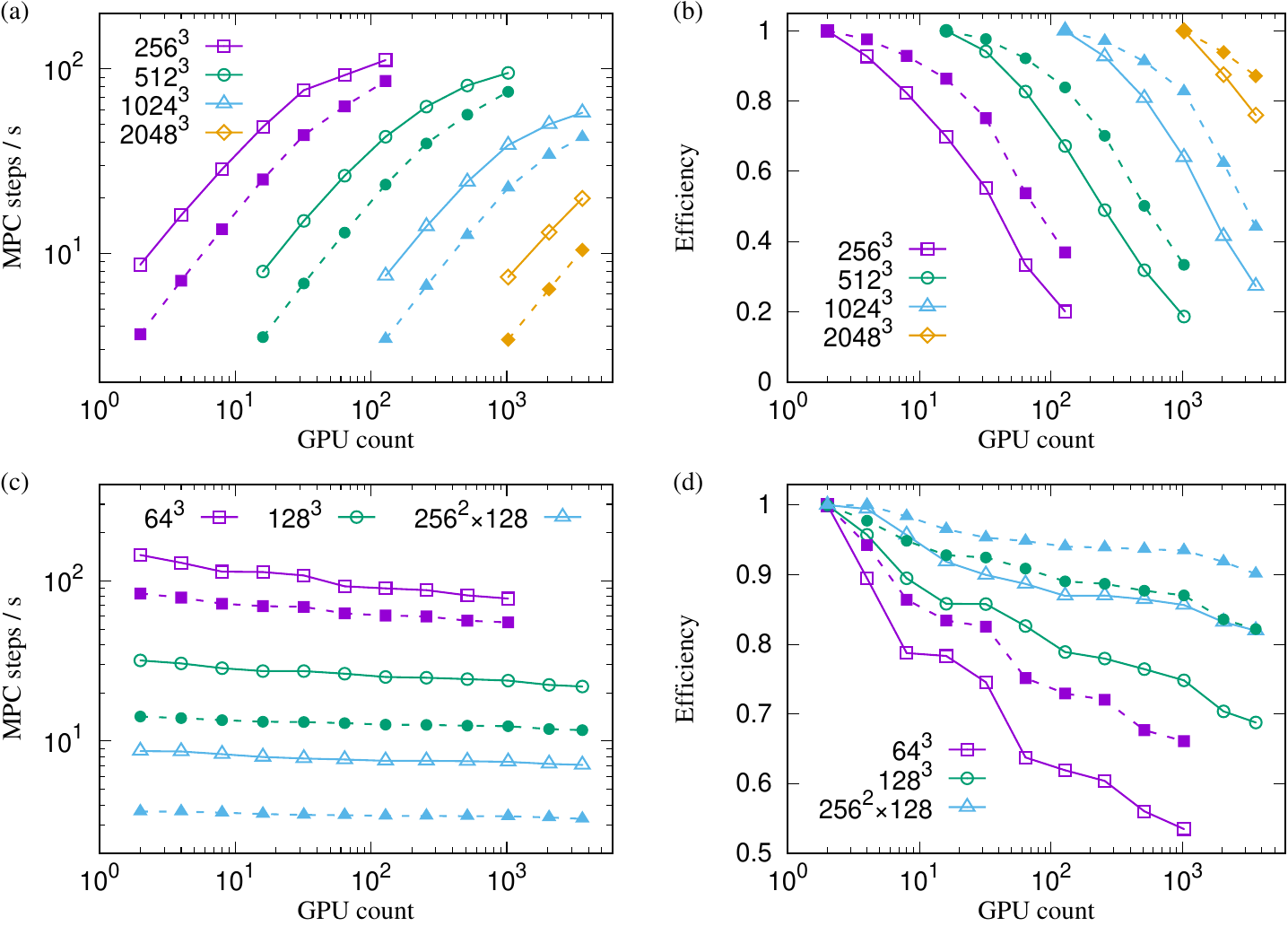}
        \caption{\label{fig:GPU_scaling}
	(a,b) Strong and (c,d) weak scaling behavior for $\langle N_c \rangle =50$. (a,c) The number of MPC steps per second,  as well as (b,d) the corresponding efficiency are shown. Filled symbols (dashed lines) and open symbols (solid lines) represent scaling behaviors of the code with and without AMC, respectively.
}
\end{figure*}

\begin{figure*}
	\centering
  	\includegraphics[width= 2\columnwidth]{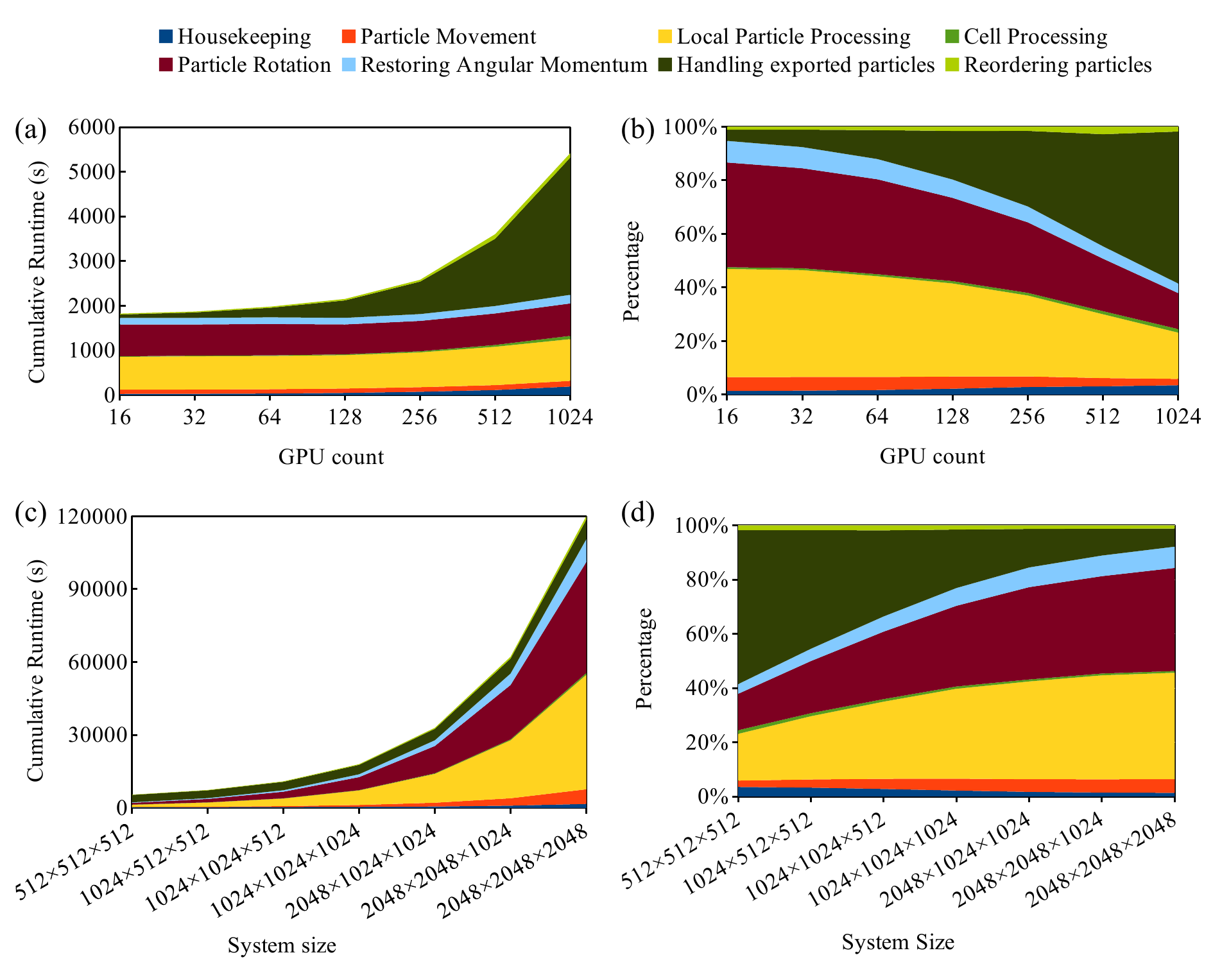}
	\caption{\label{fig:runtime}
	(a), (c) Cumulative runtime distributions for different tasks, values are summed up runtimes for all ranks, and (b), (d) the percentage of the cumulative runtime occupied by each component. While the results obtained from a fixed system size of $512\times 512\times 512$, distributed over different GPU counts as indicated, are presented in (a) and (b), the dependence on system sizes, benchmarked on 1024 GPUs, are shown in (c) and (d). In all cases, $\langle N_c \rangle = 50$, and 400 MPC steps were performed.
}
\end{figure*}

\subsubsection{Input/Output}
It is possible to store and read complete system data to/from disk. The state variables (fluid particle positions and velocities) and parameters (including density, random seed, and simulation step count) of the system are written into an easy-to-parse text file, while all particle data and select cell data is written in a binary format. Before writing data, all particles are ordered by their unshifted cell index. The cell offsets are calculated and also saved to enable reading the saved data back into systems of different rank counts or domain decompositions. All data is written to disk by using collective MPI I/O. The space necessary to store a system can be determined by multiplying the number of fluid particles with the storage sizes of their positions, velocities and IDs (if applicable) from the information in Table~\ref{tab:particle}. Additionally, for each cell the random seed, particle count and offset are saved as listed in Table~\ref{tab:cell}. As the amount of header information is negligible, for system of default precision, the estimated amount of data stored reads
\begin{align}
D	= 	&N \times [\textrm{\texttt{sizeof}(position)+\texttt{sizeof}(velocity)+\texttt{sizeof}(ID)}] \nonumber \\
		&+ N_{\rm cell}\times [\textrm{\texttt{sizeof}(random\ seed)+\texttt{sizeof}(cell offset)} \nonumber \\
  & \hspace{2cm} \textrm{+\texttt{sizeof}(particle\ count)}] \nonumber \\
	=&N \times (12+24+8) + N_{\rm cell} \times (8+8+4).
 \end{align}
For systems with $2^{32}$ or less particles, the ID tags and cell offsets are saved as 32 bit integer values requiring 4 bytes of storage space. It is also possible to generate arbitrary particle IDs for debug purposes when reading data that were saved in runs without particle IDs. 

\section{Performance/Benchmarks}

Benchmarking is performed on the JUWELS Booster System at J\"ulich Supercomputing Center~\cite{juwe:21}. Each of the nodes provides 2 AMD EPYC CPUs (24 cores each) and 4 Nvidia A100 GPUs. The nodes are connected via 4 $\times$ HDR200 Infiniband each. Benchmark test were performed utilizing up to 3540 GPUs or 12288 CPU cores for a wide range of system sizes from $64 \times 64 \times 64$ to $4096 \times 4096 \times 2048$ collision cells at low and high particle densities, i.e., $\langle N_c \rangle=$10, and 50, respectively, with and without local angular-momentum conservation. While the CPU- and GPU-based measurements for $\langle N_c \rangle=10$ are presented in SI, we focus on GPU-based measurements for $\langle N_c \rangle=50$ in this main text.

\paragraph*{\bf Strong Scaling.}
During each step, the MPC algorithm performs a relatively small number of calculations on many particles. Therefore maximizing data throughput, which can be achieved by localizing data, is the key to good performance. Accordingly, the strong scaling presented in Fig.~\ref{fig:GPU_scaling}(a) and (b) shows good efficiency as long as the workload per GPU is high. The scaling efficiency of setups with computationally more expensive options like local angular-momentum conservation or large densities (see also Fig. S1) decreases more slowly, because of their better GPU utilization. For $\langle N_c \rangle = 50$ with the AMC option, a speed-up by a factor of 10 is achieved roughly at 70\% efficiency. With a growing number of ranks, communication times become prevalent, as shown in Fig.~\ref{fig:runtime}(a) and (b). On the other hand, communication times can be kept on the order of 10\% even for 1024 GPUs for sufficiently large simulations, as shown in Fig.~\ref{fig:runtime}(c) and (d). 

\paragraph*{\bf Weak Scaling.}
Weak scaling depends on the ratio of the performance of nodes for a given load and the additional communication effort necessary for adding more notes with the same load. As shown in Fig.~\ref{fig:runtime}, the ratio of calculation versus communication for MPC on GPUs depends heavily on the utilization of the GPUs. This leads to very good weak-scaling behavior already for medium sized workloads, presented in Fig.~\ref{fig:GPU_scaling}(c) and (d). Weak scaling efficiency for large workloads is almost ideal even for very high GPU counts.

\begin{figure}
	\centering
  	\includegraphics[width= 0.94\columnwidth]{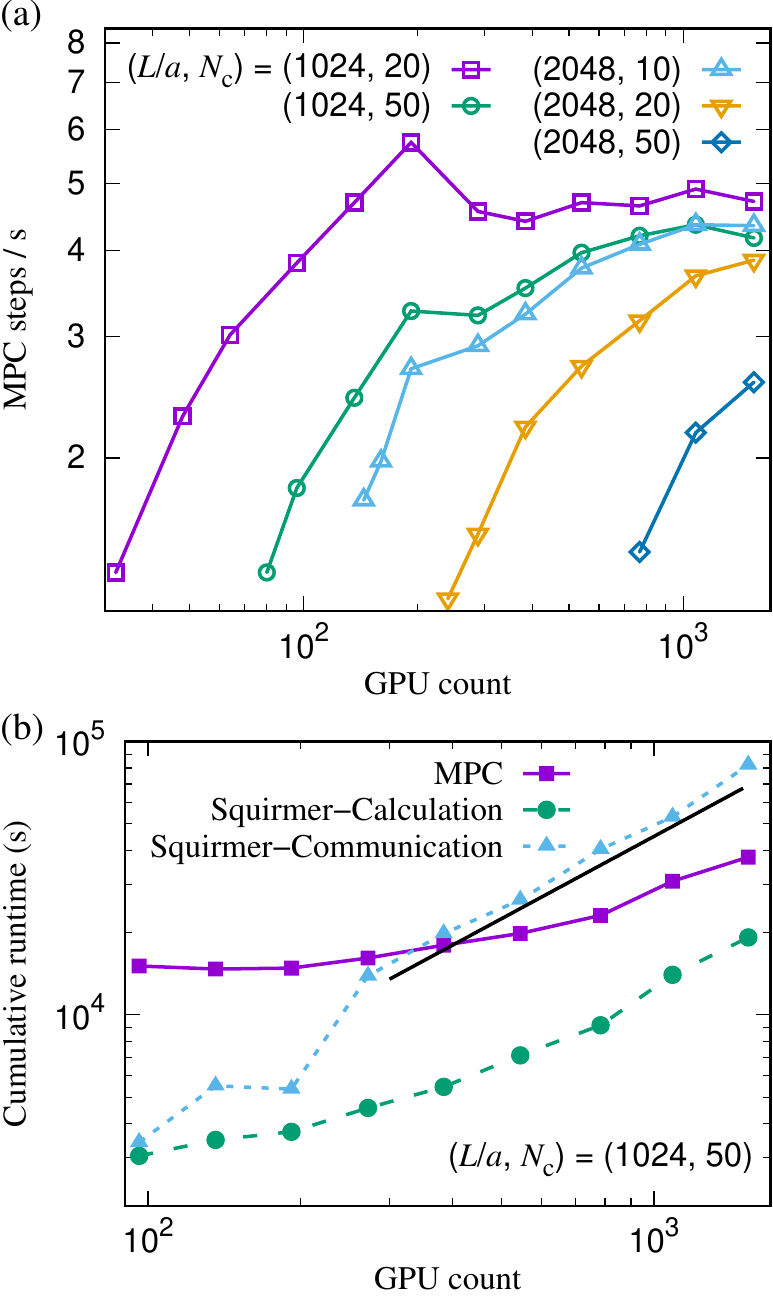}
	\caption{\label{fig:squirmer_scaling}
	(a) Scaling behaviors of the code for suspensions of $10^6$ colloidal particles with varying system sizes and MPC density. (b) Cumulative runtimes of tasks. The black solid line shows a linear scaling.
}
\end{figure}

\section{Showcases} \label{sec:showcase}
In order to demonstrate the versatility of our code, we present the results of a plugin which is capable of simulating passive and active colloidal particles in solution. The rigid-body translational dynamics of colloidal particles is governed by Newton's equations of motion. For their rotational motion, we employ the quaternion-based formulation, which allows the simulation studies of spherical and non-spherical colloids~\cite{thee:16.1} as well as chiral particles~\cite{goh:23} in a unified framework. Fluid-colloid interactions imply a linear and angular momentum transfer by solid-body interactions. We note that this momentum exchange not only affect the MPC fluid particles, but also the colloidal particles. Moreover, for active particles with surface actuation, the slip velocity ${\bm u}_{\rm slip}$ at their surface is typically non-zero. 

The implementation of colloids requires a plugin that contains multiple features described above -- such as angular-momentum conservation for proper hydrodynamics, and additional routines for boundary conditions corresponding to fluid-colloid collisions and for processing of phantom particles, in addition to simulating colloid-colloid interactions. The system can be distributed over an arbitrary number of MPI ranks and all interactions between colloids and fluid or phantom particles are processed locally within each subdomain.

\subsection{Colloidal suspensions at thermal equilibrium}
The parallel implementation for colloidal particles in a fluid can of course not be as efficient as for pure fluids, as the interactions with the fluid need to be combined and synchronized for all $N_{\rm col}$ colloids in the system. To examine the scalability quantitatively, we again perform benchmark simulations, now including colloidal particles. Here, we consider $N_{\rm col} = 10^6$ spherical colloids with effective radius $R_{\rm eff} = 3.25a$. Purely repulsive steric interaction between colloids is modeled via the separation-shifted Lennard-Jones potential, see SI, Sec.~S-IV for details. For the MPC fluid, we use the collision time $h=0.02 a\sqrt{m/(k_{\rm B}T)}$ and the rotation angle $\alpha = 130^{\circ}$ for various system sizes and MPC fluid densities.

As shown in Fig.~\ref{fig:squirmer_scaling}(a), the simulated system shows good scaling behavior of run times with the number of GPUs, with two to five times speedup, depending on the system size, beyond which saturation in the speedup is observed. Examination of cumulative run times occupied by sub-tasks, shown in Fig.~\ref{fig:squirmer_scaling}(b), indicates that the communication effort for colloid-data synchronization depends linearly on the number of MPI ranks $N_{\rm rank}$, i.e., roughly proportional to $N_{\rm col} \times N_{\rm rank}$, and therefore, becomes dominant over the efficiently distributed workload for MPC and the actual colloid calculations, which indeed exhibit sub-linear scaling.

\begin{figure}
	\centering
        \includegraphics[width= 0.95\columnwidth]{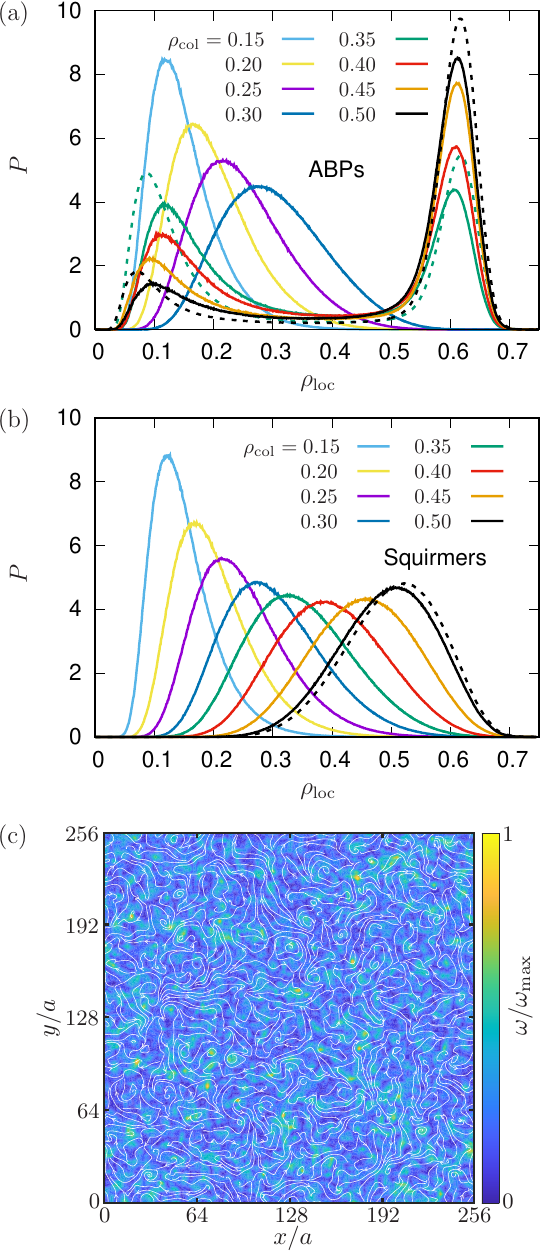}
	\caption{\label{fig:MIPS} Local density profiles for (a) ABPs, and (b) squirmers, obtained from Voronoi construction, adjusted by the volume of each Voronoi cells, see SI, Sec.~S-VI\,A for details. In both cases, ${\rm Pe}=128$. We use global densities of $\rho_{\rm col} =$ 0.15, 0.2, 0.25, 0.3, 0.35, 0.4, 0.45, and 0.5 for $L=256a$ (solid lines), and $\rho_{\rm col}=$0.35, and 0.5 for $L=512a$ (dashed lines), respectively, as indicated. 
    (c) A cross-section of the fluid velocity field (white arrow lines) and the magnitude of vorticity (heat map) are shown for squirmer suspension at $\rho_{\rm col}=0.35$. Here, a slice of the thickness $6a=2R_c$ is used and $\omega_{\rm max}\approx 1.5 \frac{u_0}{R_c} \gg D_R$.
}
\end{figure}

\subsection{Active systems of self-propelling squirmers}
As an example of active (non-equilibrium) systems, we implement a system of spherical squirmers, which model the self-propulsion of microswimmers via active surface actuation~\cite{ligh:52,blak:71,goet:10,pak:14,thee:16.1}. Specifically, we consider neutral squirmers with the swimming speed $u_0$, whose non-zero surface slip velocity is 
\begin{align}
{\bm u}_{\rm slip}^b = \frac{3}{2}u_0 \sin{\theta} \, {\bm e}_{\theta},
\end{align}
where $\theta$ denotes the polar angle in the body-fixed frame, and ${\bm e}_{\theta}$ the unit vector in the $\theta$ direction. Squirmer steric repulsion is again modeled via the Lennard-Jones potential of Eq.~(S3) in SI. We also note that the amount of additional calculations for the squirmer activity is negligible and the corresponding scaling behavior (results not shown) is almost identical to that of passive colloidal suspensions presented in Fig.~\ref{fig:squirmer_scaling}. 

We first consider a dry system (no hydrodynamic interactions) of spherical active Brownian particles (ABPs) as a reference, whose dynamics is characterized by the emergence of a motility-induced phase separation (MIPS) \cite{wyso:14,sten:14} for strong self-propulsion, or high P\'eclet number. Here, we use ABPs of radius $R_c = 3.25a$ and P\'eclet number ${\rm Pe} = u_0/(2R_c D_R) = 128$ with the rotational diffusion coefficient $D_R$ and self-propulsion speed $u_0$, and the linear system size of $L=256a$ with periodic boundary conditions in all three directions, see SI, Sec.~S-V for further simulation details. The volume packing fraction is varied from $\rho_{\rm col}=0.15$ to $0.50$, which correspond to $N_{\rm col} = 17,408$ to $N_{\rm col}=57,600$. The distribution of local particle densities is obtained via Voronoi tessellation~\cite{rycr:09} as described in Sec.~S-VI\,A of SI. As shown in Fig.~\ref{fig:MIPS}(a), the ABP system exhibits MIPS, featuring a metastable liquid-gas coexistence for high packing fractions $\rho_{\rm col} \gtrsim 0.35$, in line with the previous study~\cite{wyso:14,omar:21}. The peak densities for the liquid and gas phases are found to be $\rho_{\rm liquid} \approx 0.6$ and $\rho_{\rm gas} \approx 0.1$, respectively, also in good agreement with the previous results~\cite{omar:21}. 

We then turn to squirmer systems with the equivalent P\'eclet number, colloid densities, system sizes, and boundary conditions, to examine hydrodynamic effects on MIPS. Specifically, we use a cubic system of linear size $L = 256 a$, and the MPC fluid density (particles per collision cell) $\langle N_c \rangle = 50$, which, together with the collision time $h=0.02 a\sqrt{m/(k_{\rm B}T)}$ and the rotation angle $\alpha = 130^{\circ}$, yields the fluid viscosity of $\eta = 111.3\sqrt{mk_{\rm B}T}/a^2$~\cite{nogu:08,thee:15,goh:23}. For squirmers, we consider the radius $R_{\rm c}= 3 a$ or $R_{\rm eff} = 3.25a$ (see Sec.~S-IV, SI), and $u_0 = 0.01152\sqrt{k_{\rm B}T/m}$. This parameter set results in the rotational diffusion coefficient $D_R = 1.5 \times 10^{-5}\sqrt{k_{\rm B}T/m}/a$, Reynolds number ${\rm Re} = 0.031$, and P\'eclet number ${\rm Pe}= 128$~\cite{goh:23}. %The number of squirmers varies from $N_{\rm col} = 17,408$ ($\rho_{\rm col} \approx 0.15$) to $N_{\rm col}=57,600$ ($\rho_{\rm col} \approx 0.5$).

As shown in Fig.~\ref{fig:MIPS}(b), density profiles of hydrodynamic systems consisting of neutral squirmers always exhibit a single peak around the corresponding global packing fraction for the examined range of densities, in sharp contrast to dry ABP systems, indicating that hydrodynamic interactions suppress the motility-induced phase separation, consistent with previous results~\cite{thee:18}. We also employ a larger system size of $L=512a$ with $N_{\rm col}=325,424$ and $470,596$, corresponding to $\rho_{\rm col} \approx 0.35$ and $0.50$, respectively, to check for finite-size effects. Resultant density profiles are essentially identical to those of $L=256a$ in both cases of ABPs and squirmers, confirming that finite-size effects are not significant. We also present the velocity and vorticity fields of the MPC fluid in Fig.~\ref{fig:MIPS}(c). We observe the presence of many localized, strong vortical structures evenly distributed across the system, with the typical vorticity $\omega \approx u_0/R_c$. Corresponding squirmer configurations are also provided in Sec.~S-VI\,B of SI.

\section{Discussion}
We have presented a highly parallelized implementation of multiparticle collision dynamics for hydrodynamic simulations. The MPC algorithm is well suited for parallelization on at least two layers: data access is local and there are no long-range interactions. Therefore, the system can be domain-decomposed and distributed with little communication overhead between subdomains. Additionally, many of the operations performed on the particles are independent. Therefore, the work done in each subdomain can be further parallelized. However, we note that communication is usually much more expensive than calculation, especially for distributed systems utilizing GPUs. Nevertheless, we have confirmed overall good scaling behaviors up to more than a trillion MPC fluid particles.

Several features of the current code deviate from the previous implementations~\cite{west:14,howa:18}. The major differences can be summmarized as follows: (i) Our new implementation is plugin-based. (ii) The code provides a reproducibility option. (iii) We have abandoned the cell-list algorithm. Instead, our new version uses atomic operation directly. (iv) Every collision cell is uniquely assigned to one subdomain to avoid the necessity of partially processing additional cells in halo areas between subdomains and combining the results. (v) Instead of double buffering, we consider different variants of a memory-saving algorithm during reordering, which significantly reduces the memory footprint. 

Lastly, as a showcase, we have considered the emergence of motility-induced phase separation (MIPS) in a system of self-propelled squirmers with hydrodynamic interactions. In dry systems of active Brownian particles in three dimensions, MIPS occurs for ${\rm Pe} \gtrsim 2\textrm{-}30$~\cite{wyso:14,sten:14,omar:21}. Meanwhile, for squirmer suspensions in two dimensions,  MPC simulations demonstrate that hydrodynamic interactions significantly affect the collective behavior of active particles, and completely suppress phase separation~\cite{thee:18}. However, direct simulation for three-dimensional systems was not
achieved so far, most probably due to the computational complexity. Therefore, the results presented here clearly show that our newly implemented code is indeed very efficient, versatile and useful, facilitating investigations of large system sizes that could not be achieved so far.

\section*{PROGRAM SUMMARY}
\noindent
%{\em Program Title:}     HTMPC                                     \\
%{\em CPC Library link to program files:} (to be added by Technical Editor) \\
%{\em Developer's repository link:} TBD \\
%{\em Code Ocean capsule:} (to be added by Technical Editor)\\
%{\em Licensing provisions:} MIT  \\
{\em Programming language:} C++17, CUDA C++ (optional), MPI (optional)  \\
{\em Supplementary material:} Supplementary Information, User Manual \\
  % Fill in if necessary, otherwise leave out.
%{\em Journal reference of previous version:}*                  \\
  %Only required for a New Version summary, otherwise leave out.
%{\em Does the new version supersede the previous version?:}*   \\
  %Only required for a New Version summary, otherwise leave out.
%{\em Reasons for the new version:*}\\
  %Only required for a New Version summary, otherwise leave out.
%{\em Summary of revisions:}*\\
  %Only required for a New Version summary, otherwise leave out.
{\em Nature of problem:} Complex fluids in soft, active, and living matter are characterized by a wide range of relevant length- and time-scales, from nanometers to millimeters, and from sub-microseconds to seconds. Their dynamics is often governed by the hydrodynamics of the embedding aqueous medium. Thus, it is essential for the numerical study of such systems to develop efficient simulation techniques and highly parallel computer codes, especially when large system sizes and emergent collective behavior are considered. Several mesoscale simulation techniques have been developed in the last decades for this purpose. Multi-particle collision dynamics (MPC), a particle-based hydrodynamics simulation technique, is a promising ansatz for such an
endeavour. It is also important to develop an easy-to-extend implementation, so that the code can be adapted to various soft and living matter systems as desired. \\
  %Describe the nature of the problem here. \\
{\em Solution method:} We develop an implementation of MPC that can exploit large-scale high-performance computing resources for hydrodynamic simulations of complex fluids.
The code provides a C++ template library, which is plugin-based and can be extended by user-written plugins, implementing particles or objects interacting with the surrounding fluid. Calculations can be distributed over an arbitrary number of MPI ranks and accelerated with the current implementation supporting CUDA-capable GPUs. The code includes essential features of state-of-the-art MPC algorithms, e.g., thermostat, local angular-momentum conservation, and a variety of boundary conditions, such as periodic, no-slip (both also supporting shear flow) and slip. Simulation data can be written to and read from disk. \\
  %Describe the method solution here.
{\em Additional comments including restrictions and unusual features:}  The code provides an option to produce perfectly reproducible particle trajectories, independent of the MPI setup of a simulation system, as long as the underlying architecture and selected features are identical. It shows good scaling behaviors for sufficiently heavy workloads even for very large problems utilizing thousands of GPUs. \\
  %Provide any additional comments here.
   \\

\section*{CRediT authorship contribution statement}
%Conceptualization, Data curation, Formal Analysis, Funding acquisition, Investigation, Methodology, Project administration, Resources, Software, Supervision, Validation, Visualization, Writing – original draft, Writing – review \& editing. \\

{\bf Elmar Westphal}: Data curation, Formal Analysis, Investigation, Methodology, Software, Validation, Visualization, Writing – original draft. {\bf Segun Goh}: Data curation, Formal Analysis, Investigation, Validation, Visualization, Writing – original draft. {\bf Roland G. Winkler}: Conceptualization, Methodology, Project administration, Supervision, Writing – review \& editing. {\bf Gerhard Gompper}: Conceptualization, Funding acquisition, Methodology, Project administration, Resources, Supervision, Writing – review \& editing.

\section*{Declaration of competing interest}
The authors declare that they have no known competing financial interests or personal relationships that could have appeared to influence the work reported in this paper.

\section*{Data availability}
Data will be made available on request. 

\section*{Acknowledgement}
The authors gratefully acknowledge the Gauss Centre for Supercomputing e.V. (www.gauss-centre.eu) for funding this project by providing computing time through the John von Neumann Institute for Computing (NIC) on the GCS Supercomputer JUWELS at J\"ulich Supercomputing Centre (JSC).

%\begin{thebibliography}{0}
%\bibitem{1}Reference 1         % This list should only contain those items referenced in the                 
%\bibitem{2}Reference 2         % Program Summary section.   
%\bibitem{3}Reference 3         % Type references in text as [1], [2], etc.
                               % This list is different from the bibliography at the end of 
                               % the Long Write-Up.
%\end{thebibliography}
%* Items marked with an asterisk are only required for new versions
%of programs previously published in the CPC Program Library.\\

%\section{}
%\label{}


%% References


%\bibliography{bibliography}

\begin{thebibliography}{91}%
\makeatletter
\providecommand \@ifxundefined [1]{%
 \@ifx{#1\undefined}
}%
\providecommand \@ifnum [1]{%
 \ifnum #1\expandafter \@firstoftwo
 \else \expandafter \@secondoftwo
 \fi
}%
\providecommand \@ifx [1]{%
 \ifx #1\expandafter \@firstoftwo
 \else \expandafter \@secondoftwo
 \fi
}%
\providecommand \natexlab [1]{#1}%
\providecommand \enquote  [1]{``#1''}%
\providecommand \bibnamefont  [1]{#1}%
\providecommand \bibfnamefont [1]{#1}%
\providecommand \citenamefont [1]{#1}%
\providecommand \href@noop [0]{\@secondoftwo}%
\providecommand \href [0]{\begingroup \@sanitize@url \@href}%
\providecommand \@href[1]{\@@startlink{#1}\@@href}%
\providecommand \@@href[1]{\endgroup#1\@@endlink}%
\providecommand \@sanitize@url [0]{\catcode `\\12\catcode `\$12\catcode
  `\&12\catcode `\#12\catcode `\^12\catcode `\_12\catcode `\%12\relax}%
\providecommand \@@startlink[1]{}%
\providecommand \@@endlink[0]{}%
\providecommand \url  [0]{\begingroup\@sanitize@url \@url }%
\providecommand \@url [1]{\endgroup\@href {#1}{\urlprefix }}%
\providecommand \urlprefix  [0]{URL }%
\providecommand \Eprint [0]{\href }%
\providecommand \doibase [0]{https://doi.org/}%
\providecommand \selectlanguage [0]{\@gobble}%
\providecommand \bibinfo  [0]{\@secondoftwo}%
\providecommand \bibfield  [0]{\@secondoftwo}%
\providecommand \translation [1]{[#1]}%
\providecommand \BibitemOpen [0]{}%
\providecommand \bibitemStop [0]{}%
\providecommand \bibitemNoStop [0]{.\EOS\space}%
\providecommand \EOS [0]{\spacefactor3000\relax}%
\providecommand \BibitemShut  [1]{\csname bibitem#1\endcsname}%
\let\auto@bib@innerbib\@empty
%</preamble>
\bibitem [{\citenamefont {Gompper}\ \emph {et~al.}(2009)\citenamefont
  {Gompper}, \citenamefont {Ihle}, \citenamefont {Kroll},\ and\ \citenamefont
  {Winkler}}]{gomp:09}%
  \BibitemOpen
  \bibfield  {author} {\bibinfo {author} {\bibfnamefont {G.}~\bibnamefont
  {Gompper}}, \bibinfo {author} {\bibfnamefont {T.}~\bibnamefont {Ihle}},
  \bibinfo {author} {\bibfnamefont {D.~M.}\ \bibnamefont {Kroll}},\ and\
  \bibinfo {author} {\bibfnamefont {R.~G.}\ \bibnamefont {Winkler}},\
  }\bibfield  {title} {\bibinfo {title} {Multi-particle collision dynamics: A
  particle-based mesoscale simulation approach to the hydrodynamics of complex
  fluids},\ }\href {https://doi.org/10.1007/978-3-540-87706-6_1} {\bibfield
  {journal} {\bibinfo  {journal} {Adv. Polym. Sci.}\ }\textbf {\bibinfo
  {volume} {221}},\ \bibinfo {pages} {1} (\bibinfo {year} {2009})}\BibitemShut
  {NoStop}%
\bibitem [{\citenamefont {Dhont}\ \emph {et~al.}(2008)\citenamefont {Dhont},
  \citenamefont {Winkler}, \citenamefont {N{\"a}gele}, \citenamefont
  {Richter},\ and\ \citenamefont {Gompper}}]{dhon:08}%
  \BibitemOpen
  \bibfield  {author} {\bibinfo {author} {\bibfnamefont {J.~K.}\ \bibnamefont
  {Dhont}}, \bibinfo {author} {\bibfnamefont {R.~G.}\ \bibnamefont {Winkler}},
  \bibinfo {author} {\bibfnamefont {G.}~\bibnamefont {N{\"a}gele}}, \bibinfo
  {author} {\bibfnamefont {D.}~\bibnamefont {Richter}},\ and\ \bibinfo {author}
  {\bibfnamefont {G.}~\bibnamefont {Gompper}},\ }\href@noop {} {\emph {\bibinfo
  {title} {Soft Matter-From Synthetic to Biological Materials: Lecture Notes of
  the 39th Spring School 2008; This Spring School was organized by the
  Institute of Solid State Research in the Research Centre J{\"u}lich on 3-14
  March, 2008}}},\ \bibinfo {number} {PreJuSER-510}\ (\bibinfo  {publisher}
  {Streumethoden},\ \bibinfo {year} {2008})\BibitemShut {NoStop}%
\bibitem [{\citenamefont {Doi}\ and\ \citenamefont {Edwards}(1986)}]{doi:86}%
  \BibitemOpen
  \bibfield  {author} {\bibinfo {author} {\bibfnamefont {M.}~\bibnamefont
  {Doi}}\ and\ \bibinfo {author} {\bibfnamefont {S.~F.}\ \bibnamefont
  {Edwards}},\ }\href@noop {} {\emph {\bibinfo {title} {The Theory of Polymer
  Dynamics}}}\ (\bibinfo  {publisher} {Clarendon Press},\ \bibinfo {address}
  {Oxford},\ \bibinfo {year} {1986})\BibitemShut {NoStop}%
\bibitem [{\citenamefont {Huang}\ \emph
  {et~al.}(2010{\natexlab{a}})\citenamefont {Huang}, \citenamefont {Winkler},
  \citenamefont {Sutmann},\ and\ \citenamefont {Gompper}}]{huan:10}%
  \BibitemOpen
  \bibfield  {author} {\bibinfo {author} {\bibfnamefont {C.-C.}\ \bibnamefont
  {Huang}}, \bibinfo {author} {\bibfnamefont {R.~G.}\ \bibnamefont {Winkler}},
  \bibinfo {author} {\bibfnamefont {G.}~\bibnamefont {Sutmann}},\ and\ \bibinfo
  {author} {\bibfnamefont {G.}~\bibnamefont {Gompper}},\ }\bibfield  {title}
  {\bibinfo {title} {Semidilute polymer solutions at equilibrium and under
  shear flow},\ }\href {https://doi.org/10.1021/ma101836x} {\bibfield
  {journal} {\bibinfo  {journal} {Macromolecules}\ }\textbf {\bibinfo {volume}
  {43}},\ \bibinfo {pages} {10107} (\bibinfo {year}
  {2010}{\natexlab{a}})}\BibitemShut {NoStop}%
\bibitem [{\citenamefont {Theers}\ \emph
  {et~al.}(2016{\natexlab{a}})\citenamefont {Theers}, \citenamefont {Westphal},
  \citenamefont {Gompper},\ and\ \citenamefont {Winkler}}]{thee:16}%
  \BibitemOpen
  \bibfield  {author} {\bibinfo {author} {\bibfnamefont {M.}~\bibnamefont
  {Theers}}, \bibinfo {author} {\bibfnamefont {E.}~\bibnamefont {Westphal}},
  \bibinfo {author} {\bibfnamefont {G.}~\bibnamefont {Gompper}},\ and\ \bibinfo
  {author} {\bibfnamefont {R.~G.}\ \bibnamefont {Winkler}},\ }\bibfield
  {title} {\bibinfo {title} {From local to hydrodynamic friction in {B}rownian
  motion: A multiparticle collision dynamics simulation study},\ }\href
  {https://doi.org/10.1103/PhysRevE.93.032604} {\bibfield  {journal} {\bibinfo
  {journal} {Phys. Rev. E}\ }\textbf {\bibinfo {volume} {93}},\ \bibinfo
  {pages} {032604} (\bibinfo {year} {2016}{\natexlab{a}})}\BibitemShut
  {NoStop}%
\bibitem [{\citenamefont {Dhont}(1996)}]{dhon:96}%
  \BibitemOpen
  \bibfield  {author} {\bibinfo {author} {\bibfnamefont {J.~K.~G.}\
  \bibnamefont {Dhont}},\ }\href@noop {} {\emph {\bibinfo {title} {An
  Introduction to Dynamics of Colloids}}}\ (\bibinfo  {publisher} {Elsevier},\
  \bibinfo {address} {Amsterdam},\ \bibinfo {year} {1996})\BibitemShut
  {NoStop}%
\bibitem [{\citenamefont {Padding}\ and\ \citenamefont
  {Louis}(2006)}]{padd:06}%
  \BibitemOpen
  \bibfield  {author} {\bibinfo {author} {\bibfnamefont {J.~T.}\ \bibnamefont
  {Padding}}\ and\ \bibinfo {author} {\bibfnamefont {A.~A.}\ \bibnamefont
  {Louis}},\ }\bibfield  {title} {\bibinfo {title} {Hydrodynamic interactions
  and {Brownian} forces in colloidal suspensions: Coarse-graining over time and
  length scales},\ }\href {https://doi.org/10.1103/PhysRevE.74.031402}
  {\bibfield  {journal} {\bibinfo  {journal} {Phys. Rev. E}\ }\textbf {\bibinfo
  {volume} {74}},\ \bibinfo {pages} {031402} (\bibinfo {year}
  {2006})}\BibitemShut {NoStop}%
\bibitem [{\citenamefont {Chelakkot}\ \emph {et~al.}(2012)\citenamefont
  {Chelakkot}, \citenamefont {Winkler},\ and\ \citenamefont
  {Gompper}}]{chel:12}%
  \BibitemOpen
  \bibfield  {author} {\bibinfo {author} {\bibfnamefont {R.}~\bibnamefont
  {Chelakkot}}, \bibinfo {author} {\bibfnamefont {R.~G.}\ \bibnamefont
  {Winkler}},\ and\ \bibinfo {author} {\bibfnamefont {G.}~\bibnamefont
  {Gompper}},\ }\bibfield  {title} {\bibinfo {title} {Flow-induced helical
  coiling of semiflexible polymers in structured microchannels},\ }\href
  {https://doi.org/10.1103/PhysRevLett.109.178101} {\bibfield  {journal}
  {\bibinfo  {journal} {Phys. Rev. Lett.}\ }\textbf {\bibinfo {volume} {109}},\
  \bibinfo {pages} {178101} (\bibinfo {year} {2012})}\BibitemShut {NoStop}%
\bibitem [{\citenamefont {Wan}(2023)}]{wan:23}%
  \BibitemOpen
  \bibfield  {author} {\bibinfo {author} {\bibfnamefont {K.~Y.}\ \bibnamefont
  {Wan}},\ }\bibfield  {title} {\bibinfo {title} {Life through the fluid
  dynamics lens},\ }\href {https://doi.org/10.1038/s41567-023-02299-7}
  {\bibfield  {journal} {\bibinfo  {journal} {Nat. Phys.}\ }\textbf {\bibinfo
  {volume} {19}},\ \bibinfo {pages} {1744} (\bibinfo {year}
  {2023})}\BibitemShut {NoStop}%
\bibitem [{\citenamefont {Noguchi}\ and\ \citenamefont
  {Gompper}(2005)}]{nogu:05}%
  \BibitemOpen
  \bibfield  {author} {\bibinfo {author} {\bibfnamefont {H.}~\bibnamefont
  {Noguchi}}\ and\ \bibinfo {author} {\bibfnamefont {G.}~\bibnamefont
  {Gompper}},\ }\bibfield  {title} {\bibinfo {title} {Shape transitions of
  fluid vesicles and red blood cells in capillary flow},\ }\href
  {https://doi.org/10.1073/pnas.0504243102} {\bibfield  {journal} {\bibinfo
  {journal} {Proc. Natl. Acad. Sci. USA}\ }\textbf {\bibinfo {volume} {102}},\
  \bibinfo {pages} {14159} (\bibinfo {year} {2005})}\BibitemShut {NoStop}%
\bibitem [{\citenamefont {Fedosov}\ \emph {et~al.}(2010)\citenamefont
  {Fedosov}, \citenamefont {Caswell},\ and\ \citenamefont
  {Karniadakis}}]{fedo:10.1}%
  \BibitemOpen
  \bibfield  {author} {\bibinfo {author} {\bibfnamefont {D.~A.}\ \bibnamefont
  {Fedosov}}, \bibinfo {author} {\bibfnamefont {B.}~\bibnamefont {Caswell}},\
  and\ \bibinfo {author} {\bibfnamefont {G.~E.}\ \bibnamefont {Karniadakis}},\
  }\bibfield  {title} {\bibinfo {title} {A multiscale red blood cell model with
  accurate mechanics, rheology, and dynamics},\ }\href
  {https://doi.org/10.1016/j.bpj.2010.02.002} {\bibfield  {journal} {\bibinfo
  {journal} {Biophys. J.}\ }\textbf {\bibinfo {volume} {98}},\ \bibinfo {pages}
  {2215} (\bibinfo {year} {2010})}\BibitemShut {NoStop}%
\bibitem [{\citenamefont {Lauga}\ and\ \citenamefont {Powers}(2009)}]{laug:09}%
  \BibitemOpen
  \bibfield  {author} {\bibinfo {author} {\bibfnamefont {E.}~\bibnamefont
  {Lauga}}\ and\ \bibinfo {author} {\bibfnamefont {T.~R.}\ \bibnamefont
  {Powers}},\ }\bibfield  {title} {\bibinfo {title} {{T}he hydrodynamics of
  swimming microorganisms},\ }\href
  {https://doi.org/10.1088/0034-4885/72/9/096601} {\bibfield  {journal}
  {\bibinfo  {journal} {Rep. Prog. Phys.}\ }\textbf {\bibinfo {volume} {72}},\
  \bibinfo {pages} {096601} (\bibinfo {year} {2009})}\BibitemShut {NoStop}%
\bibitem [{\citenamefont {Gompper}\ \emph {et~al.}(2016)\citenamefont
  {Gompper}, \citenamefont {Bechinger}, \citenamefont {Herminghaus},
  \citenamefont {Isele-Holder}, \citenamefont {Kaupp}, \citenamefont
  {L{\"o}wen}, \citenamefont {Stark},\ and\ \citenamefont {Winkler}}]{gomp:16}%
  \BibitemOpen
  \bibfield  {author} {\bibinfo {author} {\bibfnamefont {G.}~\bibnamefont
  {Gompper}}, \bibinfo {author} {\bibfnamefont {C.}~\bibnamefont {Bechinger}},
  \bibinfo {author} {\bibfnamefont {S.}~\bibnamefont {Herminghaus}}, \bibinfo
  {author} {\bibfnamefont {R.}~\bibnamefont {Isele-Holder}}, \bibinfo {author}
  {\bibfnamefont {U.~B.}\ \bibnamefont {Kaupp}}, \bibinfo {author}
  {\bibfnamefont {H.}~\bibnamefont {L{\"o}wen}}, \bibinfo {author}
  {\bibfnamefont {H.}~\bibnamefont {Stark}},\ and\ \bibinfo {author}
  {\bibfnamefont {R.~G.}\ \bibnamefont {Winkler}},\ }\bibfield  {title}
  {\bibinfo {title} {Microswimmers--from single particle motion to collective
  behavior},\ }\href {https://doi.org/10.1140/epjst/e2016-60095-3} {\bibfield
  {journal} {\bibinfo  {journal} {Eur. Phys. J. Spec. Top.}\ }\textbf {\bibinfo
  {volume} {225}},\ \bibinfo {pages} {2061} (\bibinfo {year}
  {2016})}\BibitemShut {NoStop}%
\bibitem [{\citenamefont {Hu}\ \emph {et~al.}(2015{\natexlab{a}})\citenamefont
  {Hu}, \citenamefont {Wysocki}, \citenamefont {Winkler},\ and\ \citenamefont
  {Gompper}}]{hu:15}%
  \BibitemOpen
  \bibfield  {author} {\bibinfo {author} {\bibfnamefont {J.}~\bibnamefont
  {Hu}}, \bibinfo {author} {\bibfnamefont {A.}~\bibnamefont {Wysocki}},
  \bibinfo {author} {\bibfnamefont {R.~G.}\ \bibnamefont {Winkler}},\ and\
  \bibinfo {author} {\bibfnamefont {G.}~\bibnamefont {Gompper}},\ }\bibfield
  {title} {\bibinfo {title} {Physical sensing of surface properties by
  microswimmers -- directing bacterial motion via wall slip},\ }\href
  {https://doi.org/10.1038/srep09586} {\bibfield  {journal} {\bibinfo
  {journal} {Sci. Rep.}\ }\textbf {\bibinfo {volume} {5}},\ \bibinfo {pages}
  {9586} (\bibinfo {year} {2015}{\natexlab{a}})}\BibitemShut {NoStop}%
\bibitem [{\citenamefont {Mousavi}\ \emph {et~al.}(2020)\citenamefont
  {Mousavi}, \citenamefont {Gompper},\ and\ \citenamefont {Winkler}}]{mous:20}%
  \BibitemOpen
  \bibfield  {author} {\bibinfo {author} {\bibfnamefont {S.~M.}\ \bibnamefont
  {Mousavi}}, \bibinfo {author} {\bibfnamefont {G.}~\bibnamefont {Gompper}},\
  and\ \bibinfo {author} {\bibfnamefont {R.~G.}\ \bibnamefont {Winkler}},\
  }\bibfield  {title} {\bibinfo {title} {Wall entrapment of peritrichous
  bacteria: a mesoscale hydrodynamics simulation study},\ }\href
  {https://doi.org/10.1039/D0SM00571A} {\bibfield  {journal} {\bibinfo
  {journal} {Soft Matter}\ }\textbf {\bibinfo {volume} {16}},\ \bibinfo {pages}
  {4866} (\bibinfo {year} {2020})}\BibitemShut {NoStop}%
\bibitem [{\citenamefont {Elgeti}\ \emph {et~al.}(2015)\citenamefont {Elgeti},
  \citenamefont {Winkler},\ and\ \citenamefont {Gompper}}]{elge:15}%
  \BibitemOpen
  \bibfield  {author} {\bibinfo {author} {\bibfnamefont {J.}~\bibnamefont
  {Elgeti}}, \bibinfo {author} {\bibfnamefont {R.~G.}\ \bibnamefont
  {Winkler}},\ and\ \bibinfo {author} {\bibfnamefont {G.}~\bibnamefont
  {Gompper}},\ }\bibfield  {title} {\bibinfo {title} {Physics of
  microswimmers---single particle motion and collective behavior: a review},\
  }\href {https://doi.org/10.1088/0034-4885/78/5/056601} {\bibfield  {journal}
  {\bibinfo  {journal} {Rep. Prog. Phys.}\ }\textbf {\bibinfo {volume} {78}},\
  \bibinfo {pages} {056601} (\bibinfo {year} {2015})}\BibitemShut {NoStop}%
\bibitem [{\citenamefont {Hoogerbrugge}\ and\ \citenamefont
  {Koelman}(1992)}]{hoog:92}%
  \BibitemOpen
  \bibfield  {author} {\bibinfo {author} {\bibfnamefont {P.~J.}\ \bibnamefont
  {Hoogerbrugge}}\ and\ \bibinfo {author} {\bibfnamefont {J.~M. V.~A.}\
  \bibnamefont {Koelman}},\ }\bibfield  {title} {\bibinfo {title} {Simulating
  microscopic hydrodynamics phenomena with dissipative particle dynamics},\
  }\href {https://doi.org/10.1209/0295-5075/19/3/001} {\bibfield  {journal}
  {\bibinfo  {journal} {EPL}\ }\textbf {\bibinfo {volume} {19}},\ \bibinfo
  {pages} {155} (\bibinfo {year} {1992})}\BibitemShut {NoStop}%
\bibitem [{\citenamefont {Español}\ and\ \citenamefont
  {Warren}(2017)}]{espa:17}%
  \BibitemOpen
  \bibfield  {author} {\bibinfo {author} {\bibfnamefont {P.}~\bibnamefont
  {Español}}\ and\ \bibinfo {author} {\bibfnamefont {P.~B.}\ \bibnamefont
  {Warren}},\ }\bibfield  {title} {\bibinfo {title} {{Perspective: Dissipative
  particle dynamics}},\ }\href {https://doi.org/10.1063/1.4979514} {\bibfield
  {journal} {\bibinfo  {journal} {J. Chem. Phys.}\ }\textbf {\bibinfo {volume}
  {146}},\ \bibinfo {pages} {150901} (\bibinfo {year} {2017})}\BibitemShut
  {NoStop}%
\bibitem [{\citenamefont {McNamara}\ and\ \citenamefont
  {Zanetti}(1988)}]{mcna:88}%
  \BibitemOpen
  \bibfield  {author} {\bibinfo {author} {\bibfnamefont {G.~R.}\ \bibnamefont
  {McNamara}}\ and\ \bibinfo {author} {\bibfnamefont {G.}~\bibnamefont
  {Zanetti}},\ }\bibfield  {title} {\bibinfo {title} {Use of the {Boltzmann}
  equation to simulate lattice-gas automata},\ }\href
  {https://doi.org/10.1103/PhysRevLett.61.2332} {\bibfield  {journal} {\bibinfo
   {journal} {Phys. Rev. Lett.}\ }\textbf {\bibinfo {volume} {61}},\ \bibinfo
  {pages} {2332} (\bibinfo {year} {1988})}\BibitemShut {NoStop}%
\bibitem [{\citenamefont {Succi}(2001)}]{succ:01}%
  \BibitemOpen
  \bibfield  {author} {\bibinfo {author} {\bibfnamefont {S.}~\bibnamefont
  {Succi}},\ }\href@noop {} {\emph {\bibinfo {title} {The lattice Boltzmann
  equation: for fluid dynamics and beyond}}}\ (\bibinfo  {publisher} {Oxford
  University Press},\ \bibinfo {year} {2001})\BibitemShut {NoStop}%
\bibitem [{\citenamefont {D{\"u}nweg}\ and\ \citenamefont
  {Ladd}(2009)}]{duen:09}%
  \BibitemOpen
  \bibfield  {author} {\bibinfo {author} {\bibfnamefont {B.}~\bibnamefont
  {D{\"u}nweg}}\ and\ \bibinfo {author} {\bibfnamefont {A.~C.}\ \bibnamefont
  {Ladd}},\ }\bibfield  {title} {\bibinfo {title} {Lattice {B}oltzmann
  simulations of soft matter systems},\ }\href
  {https://doi.org/10.1007/978-3-540-87706-6{\_}2} {\bibfield  {journal}
  {\bibinfo  {journal} {Adv. Polym. Sci.}\ }\textbf {\bibinfo {volume} {221}},\
  \bibinfo {pages} {89} (\bibinfo {year} {2009})}\BibitemShut {NoStop}%
\bibitem [{\citenamefont {Bird}(1994)}]{bird:94}%
  \BibitemOpen
  \bibfield  {author} {\bibinfo {author} {\bibfnamefont {G.~A.}\ \bibnamefont
  {Bird}},\ }\href@noop {} {\emph {\bibinfo {title} {Molecular Gas Dynamics and
  the Direct Simulation of Gas Flows}}}\ (\bibinfo  {publisher} {Oxford
  University Press, Oxford},\ \bibinfo {year} {1994})\BibitemShut {NoStop}%
\bibitem [{\citenamefont {Plimpton}\ \emph {et~al.}(2019)\citenamefont
  {Plimpton}, \citenamefont {Moore}, \citenamefont {Borner}, \citenamefont
  {Stagg}, \citenamefont {Koehler}, \citenamefont {Torczynski},\ and\
  \citenamefont {Gallis}}]{plim:19}%
  \BibitemOpen
  \bibfield  {author} {\bibinfo {author} {\bibfnamefont {S.~J.}\ \bibnamefont
  {Plimpton}}, \bibinfo {author} {\bibfnamefont {S.~G.}\ \bibnamefont {Moore}},
  \bibinfo {author} {\bibfnamefont {A.}~\bibnamefont {Borner}}, \bibinfo
  {author} {\bibfnamefont {A.~K.}\ \bibnamefont {Stagg}}, \bibinfo {author}
  {\bibfnamefont {T.~P.}\ \bibnamefont {Koehler}}, \bibinfo {author}
  {\bibfnamefont {J.~R.}\ \bibnamefont {Torczynski}},\ and\ \bibinfo {author}
  {\bibfnamefont {M.~A.}\ \bibnamefont {Gallis}},\ }\bibfield  {title}
  {\bibinfo {title} {{Direct simulation Monte Carlo on petaflop supercomputers
  and beyond}},\ }\href {https://doi.org/10.1063/1.5108534} {\bibfield
  {journal} {\bibinfo  {journal} {Phys. Fluids}\ }\textbf {\bibinfo {volume}
  {31}},\ \bibinfo {pages} {086101} (\bibinfo {year} {2019})}\BibitemShut
  {NoStop}%
\bibitem [{\citenamefont {Kapral}(2008)}]{kapr:08}%
  \BibitemOpen
  \bibfield  {author} {\bibinfo {author} {\bibfnamefont {R.}~\bibnamefont
  {Kapral}},\ }\bibfield  {title} {\bibinfo {title} {Multiparticle collision
  dynamics: Simulations of complex systems on mesoscale},\ }\href
  {https://doi.org/10.1002/9780470371572.ch2} {\bibfield  {journal} {\bibinfo
  {journal} {Adv. Chem. Phys.}\ }\textbf {\bibinfo {volume} {140}},\ \bibinfo
  {pages} {89} (\bibinfo {year} {2008})}\BibitemShut {NoStop}%
\bibitem [{\citenamefont {Westphal}\ \emph {et~al.}(2014)\citenamefont
  {Westphal}, \citenamefont {Singh}, \citenamefont {Huang}, \citenamefont
  {Gompper},\ and\ \citenamefont {Winkler}}]{west:14}%
  \BibitemOpen
  \bibfield  {author} {\bibinfo {author} {\bibfnamefont {E.}~\bibnamefont
  {Westphal}}, \bibinfo {author} {\bibfnamefont {S.~P.}\ \bibnamefont {Singh}},
  \bibinfo {author} {\bibfnamefont {C.-C.}\ \bibnamefont {Huang}}, \bibinfo
  {author} {\bibfnamefont {G.}~\bibnamefont {Gompper}},\ and\ \bibinfo {author}
  {\bibfnamefont {R.~G.}\ \bibnamefont {Winkler}},\ }\bibfield  {title}
  {\bibinfo {title} {Multiparticle collision dynamics: {GPU} accelerated
  particle-based mesoscale hydrodynamic simulations},\ }\href
  {https://doi.org/10.1016/j.cpc.2013.10.004} {\bibfield  {journal} {\bibinfo
  {journal} {Comput. Phys. Comm.}\ }\textbf {\bibinfo {volume} {185}},\
  \bibinfo {pages} {495} (\bibinfo {year} {2014})}\BibitemShut {NoStop}%
\bibitem [{\citenamefont {Howard}\ \emph {et~al.}(2018)\citenamefont {Howard},
  \citenamefont {Panagiotopoulos},\ and\ \citenamefont
  {Nikoubashman}}]{howa:18}%
  \BibitemOpen
  \bibfield  {author} {\bibinfo {author} {\bibfnamefont {M.~P.}\ \bibnamefont
  {Howard}}, \bibinfo {author} {\bibfnamefont {A.~Z.}\ \bibnamefont
  {Panagiotopoulos}},\ and\ \bibinfo {author} {\bibfnamefont {A.}~\bibnamefont
  {Nikoubashman}},\ }\bibfield  {title} {\bibinfo {title} {Efficient mesoscale
  hydrodynamics: Multiparticle collision dynamics with massively parallel {GPU}
  acceleration},\ }\href {https://doi.org/10.1016/j.cpc.2018.04.009} {\bibfield
   {journal} {\bibinfo  {journal} {Comput. Phys. Commun.}\ }\textbf {\bibinfo
  {volume} {230}},\ \bibinfo {pages} {10} (\bibinfo {year} {2018})}\BibitemShut
  {NoStop}%
\bibitem [{\citenamefont {Huang}\ \emph {et~al.}(2012)\citenamefont {Huang},
  \citenamefont {Gompper},\ and\ \citenamefont {Winkler}}]{huan:12}%
  \BibitemOpen
  \bibfield  {author} {\bibinfo {author} {\bibfnamefont {C.-C.}\ \bibnamefont
  {Huang}}, \bibinfo {author} {\bibfnamefont {G.}~\bibnamefont {Gompper}},\
  and\ \bibinfo {author} {\bibfnamefont {R.~G.}\ \bibnamefont {Winkler}},\
  }\bibfield  {title} {\bibinfo {title} {Hydrodynamic correlations in
  multiparticle collision dynamics fluids},\ }\href
  {https://doi.org/10.1103/PhysRevE.86.056711} {\bibfield  {journal} {\bibinfo
  {journal} {Phys. Rev. E}\ }\textbf {\bibinfo {volume} {86}},\ \bibinfo
  {pages} {056711} (\bibinfo {year} {2012})}\BibitemShut {NoStop}%
\bibitem [{\citenamefont {Malevanets}\ and\ \citenamefont
  {Kapral}(1999)}]{male:99}%
  \BibitemOpen
  \bibfield  {author} {\bibinfo {author} {\bibfnamefont {A.}~\bibnamefont
  {Malevanets}}\ and\ \bibinfo {author} {\bibfnamefont {R.}~\bibnamefont
  {Kapral}},\ }\bibfield  {title} {\bibinfo {title} {Mesoscopic model for
  solvent dynamics},\ }\href {https://doi.org/10.1063/1.478857} {\bibfield
  {journal} {\bibinfo  {journal} {J. Chem. Phys.}\ }\textbf {\bibinfo {volume}
  {110}},\ \bibinfo {pages} {8605} (\bibinfo {year} {1999})}\BibitemShut
  {NoStop}%
\bibitem [{\citenamefont {Ihle}\ and\ \citenamefont
  {Kroll}(2003{\natexlab{a}})}]{ihle:03}%
  \BibitemOpen
  \bibfield  {author} {\bibinfo {author} {\bibfnamefont {T.}~\bibnamefont
  {Ihle}}\ and\ \bibinfo {author} {\bibfnamefont {D.~M.}\ \bibnamefont
  {Kroll}},\ }\bibfield  {title} {\bibinfo {title} {Stochastic rotation
  dynamics {I}: Formalism, {Galilean} invariance, {Green-Kubo} relations},\
  }\href {https://doi.org/10.1103/PhysRevE.67.066705} {\bibfield  {journal}
  {\bibinfo  {journal} {Phys. Rev. E}\ }\textbf {\bibinfo {volume} {67}},\
  \bibinfo {pages} {066705} (\bibinfo {year} {2003}{\natexlab{a}})}\BibitemShut
  {NoStop}%
\bibitem [{\citenamefont {Ihle}\ and\ \citenamefont
  {Kroll}(2003{\natexlab{b}})}]{ihle:03.1}%
  \BibitemOpen
  \bibfield  {author} {\bibinfo {author} {\bibfnamefont {T.}~\bibnamefont
  {Ihle}}\ and\ \bibinfo {author} {\bibfnamefont {D.~M.}\ \bibnamefont
  {Kroll}},\ }\bibfield  {title} {\bibinfo {title} {Stochastic rotation
  dynamics {II}: Transport coefficients, numerics, long time tails},\ }\href
  {https://doi.org/10.1103/PhysRevE.67.066706} {\bibfield  {journal} {\bibinfo
  {journal} {Phys. Rev. E}\ }\textbf {\bibinfo {volume} {67}},\ \bibinfo
  {pages} {066706} (\bibinfo {year} {2003}{\natexlab{b}})}\BibitemShut
  {NoStop}%
\bibitem [{\citenamefont {Kikuchi}\ \emph {et~al.}(2003)\citenamefont
  {Kikuchi}, \citenamefont {Pooley}, \citenamefont {Ryder},\ and\ \citenamefont
  {Yeomans}}]{kiku:03}%
  \BibitemOpen
  \bibfield  {author} {\bibinfo {author} {\bibfnamefont {N.}~\bibnamefont
  {Kikuchi}}, \bibinfo {author} {\bibfnamefont {C.~M.}\ \bibnamefont {Pooley}},
  \bibinfo {author} {\bibfnamefont {J.~F.}\ \bibnamefont {Ryder}},\ and\
  \bibinfo {author} {\bibfnamefont {J.~M.}\ \bibnamefont {Yeomans}},\
  }\bibfield  {title} {\bibinfo {title} {Transport coefficients of a mesoscopic
  fluid dynamics model},\ }\href {https://doi.org/10.1063/1.1603721} {\bibfield
   {journal} {\bibinfo  {journal} {J. Chem. Phys.}\ }\textbf {\bibinfo {volume}
  {119}},\ \bibinfo {pages} {6388} (\bibinfo {year} {2003})}\BibitemShut
  {NoStop}%
\bibitem [{\citenamefont {Noguchi}\ and\ \citenamefont
  {Gompper}(2008)}]{nogu:08}%
  \BibitemOpen
  \bibfield  {author} {\bibinfo {author} {\bibfnamefont {H.}~\bibnamefont
  {Noguchi}}\ and\ \bibinfo {author} {\bibfnamefont {G.}~\bibnamefont
  {Gompper}},\ }\bibfield  {title} {\bibinfo {title} {Transport coefficients of
  off-lattice mesoscale-hydrodynamics simulation techniques},\ }\href
  {https://doi.org/10.1103/PhysRevE.78.016706} {\bibfield  {journal} {\bibinfo
  {journal} {Phys. Rev. E}\ }\textbf {\bibinfo {volume} {78}},\ \bibinfo
  {pages} {016706} (\bibinfo {year} {2008})}\BibitemShut {NoStop}%
\bibitem [{\citenamefont {Winkler}\ and\ \citenamefont
  {Huang}(2009)}]{wink:09}%
  \BibitemOpen
  \bibfield  {author} {\bibinfo {author} {\bibfnamefont {R.~G.}\ \bibnamefont
  {Winkler}}\ and\ \bibinfo {author} {\bibfnamefont {C.-C.}\ \bibnamefont
  {Huang}},\ }\bibfield  {title} {\bibinfo {title} {Stress tensors of
  multiparticle collision dynamics fluids},\ }\href
  {https://doi.org/10.1063/1.3077860} {\bibfield  {journal} {\bibinfo
  {journal} {J. Chem. Phys.}\ }\textbf {\bibinfo {volume} {130}},\ \bibinfo
  {pages} {074907} (\bibinfo {year} {2009})}\BibitemShut {NoStop}%
\bibitem [{\citenamefont {Padding}\ and\ \citenamefont
  {Louis}(2004)}]{padd:04}%
  \BibitemOpen
  \bibfield  {author} {\bibinfo {author} {\bibfnamefont {J.~T.}\ \bibnamefont
  {Padding}}\ and\ \bibinfo {author} {\bibfnamefont {A.~A.}\ \bibnamefont
  {Louis}},\ }\bibfield  {title} {\bibinfo {title} {Hydrodynamic and {Brownian}
  fluctuations in sedimenting suspensions},\ }\href
  {https://doi.org/10.1103/PhysRevLett.93.220601} {\bibfield  {journal}
  {\bibinfo  {journal} {Phys. Rev. Lett.}\ }\textbf {\bibinfo {volume} {93}},\
  \bibinfo {pages} {220601} (\bibinfo {year} {2004})}\BibitemShut {NoStop}%
\bibitem [{\citenamefont {Hecht}\ \emph {et~al.}(2006)\citenamefont {Hecht},
  \citenamefont {Harting}, \citenamefont {Bier}, \citenamefont {Reinshagen},\
  and\ \citenamefont {Herrmann}}]{hech:06}%
  \BibitemOpen
  \bibfield  {author} {\bibinfo {author} {\bibfnamefont {M.}~\bibnamefont
  {Hecht}}, \bibinfo {author} {\bibfnamefont {J.}~\bibnamefont {Harting}},
  \bibinfo {author} {\bibfnamefont {M.}~\bibnamefont {Bier}}, \bibinfo {author}
  {\bibfnamefont {J.}~\bibnamefont {Reinshagen}},\ and\ \bibinfo {author}
  {\bibfnamefont {H.~J.}\ \bibnamefont {Herrmann}},\ }\bibfield  {title}
  {\bibinfo {title} {Shear viscosity of claylike colloids in computer
  simulations and experiments},\ }\href
  {https://doi.org/10.1103/PhysRevE.74.021403} {\bibfield  {journal} {\bibinfo
  {journal} {Phys. Rev. E}\ }\textbf {\bibinfo {volume} {74}},\ \bibinfo
  {pages} {021403} (\bibinfo {year} {2006})}\BibitemShut {NoStop}%
\bibitem [{\citenamefont {Frank}\ and\ \citenamefont
  {Winkler}(2008)}]{fran:08}%
  \BibitemOpen
  \bibfield  {author} {\bibinfo {author} {\bibfnamefont {S.}~\bibnamefont
  {Frank}}\ and\ \bibinfo {author} {\bibfnamefont {R.~G.}\ \bibnamefont
  {Winkler}},\ }\bibfield  {title} {\bibinfo {title} {Polyelectrolyte
  electrophoresis: Field effects and hydrodynamic interactions},\ }\href
  {https://doi.org/10.1209/0295-5075/83/38004} {\bibfield  {journal} {\bibinfo
  {journal} {EPL}\ }\textbf {\bibinfo {volume} {83}},\ \bibinfo {pages} {38004}
  (\bibinfo {year} {2008})}\BibitemShut {NoStop}%
\bibitem [{\citenamefont {Mandal}\ and\ \citenamefont {Mazza}(2019)}]{mand:19}%
  \BibitemOpen
  \bibfield  {author} {\bibinfo {author} {\bibfnamefont {S.}~\bibnamefont
  {Mandal}}\ and\ \bibinfo {author} {\bibfnamefont {M.~G.}\ \bibnamefont
  {Mazza}},\ }\bibfield  {title} {\bibinfo {title} {Multiparticle collision
  dynamics for tensorial nematodynamics},\ }\href
  {https://doi.org/10.1103/PhysRevE.99.063319} {\bibfield  {journal} {\bibinfo
  {journal} {Phys. Rev. E}\ }\textbf {\bibinfo {volume} {99}},\ \bibinfo
  {pages} {063319} (\bibinfo {year} {2019})}\BibitemShut {NoStop}%
\bibitem [{\citenamefont {Malevanets}\ and\ \citenamefont
  {Kapral}(2000)}]{male:00}%
  \BibitemOpen
  \bibfield  {author} {\bibinfo {author} {\bibfnamefont {A.}~\bibnamefont
  {Malevanets}}\ and\ \bibinfo {author} {\bibfnamefont {R.}~\bibnamefont
  {Kapral}},\ }\bibfield  {title} {\bibinfo {title} {Solute molecular dynamics
  in a mesoscopic solvent},\ }\href {https://doi.org/10.1063/1.481289}
  {\bibfield  {journal} {\bibinfo  {journal} {J. Chem. Phys.}\ }\textbf
  {\bibinfo {volume} {112}},\ \bibinfo {pages} {7260} (\bibinfo {year}
  {2000})}\BibitemShut {NoStop}%
\bibitem [{\citenamefont {Mussawisade}\ \emph {et~al.}(2005)\citenamefont
  {Mussawisade}, \citenamefont {Ripoll}, \citenamefont {Winkler},\ and\
  \citenamefont {Gompper}}]{muss:05}%
  \BibitemOpen
  \bibfield  {author} {\bibinfo {author} {\bibfnamefont {K.}~\bibnamefont
  {Mussawisade}}, \bibinfo {author} {\bibfnamefont {M.}~\bibnamefont {Ripoll}},
  \bibinfo {author} {\bibfnamefont {R.~G.}\ \bibnamefont {Winkler}},\ and\
  \bibinfo {author} {\bibfnamefont {G.}~\bibnamefont {Gompper}},\ }\bibfield
  {title} {\bibinfo {title} {Dynamics of polymers in a particle based
  mesoscopic solvent},\ }\href {https://doi.org/10.1063/1.2041527} {\bibfield
  {journal} {\bibinfo  {journal} {J. Chem. Phys.}\ }\textbf {\bibinfo {volume}
  {123}},\ \bibinfo {pages} {144905} (\bibinfo {year} {2005})}\BibitemShut
  {NoStop}%
\bibitem [{\citenamefont {Ali}\ \emph {et~al.}(2006)\citenamefont {Ali},
  \citenamefont {Marenduzzo},\ and\ \citenamefont {Yeomans}}]{ali:06}%
  \BibitemOpen
  \bibfield  {author} {\bibinfo {author} {\bibfnamefont {I.}~\bibnamefont
  {Ali}}, \bibinfo {author} {\bibfnamefont {D.}~\bibnamefont {Marenduzzo}},\
  and\ \bibinfo {author} {\bibfnamefont {J.~M.}\ \bibnamefont {Yeomans}},\
  }\bibfield  {title} {\bibinfo {title} {Polymer packaging and ejection in
  viral capsids: Shape matters},\ }\href
  {https://doi.org/10.1103/PhysRevLett.96.208102} {\bibfield  {journal}
  {\bibinfo  {journal} {Phys. Rev. Lett.}\ }\textbf {\bibinfo {volume} {96}},\
  \bibinfo {pages} {208102} (\bibinfo {year} {2006})}\BibitemShut {NoStop}%
\bibitem [{\citenamefont {Nikoubashman}\ and\ \citenamefont
  {Likos}(2010)}]{niko:10}%
  \BibitemOpen
  \bibfield  {author} {\bibinfo {author} {\bibfnamefont {A.}~\bibnamefont
  {Nikoubashman}}\ and\ \bibinfo {author} {\bibfnamefont {C.~N.}\ \bibnamefont
  {Likos}},\ }\bibfield  {title} {\bibinfo {title} {Flow-induced polymer
  translocation through narrow and patterned channels},\ }\href
  {https://doi.org/10.1063/1.3466918} {\bibfield  {journal} {\bibinfo
  {journal} {J. Chem. Phys.}\ }\textbf {\bibinfo {volume} {133}},\ \bibinfo
  {pages} {074901} (\bibinfo {year} {2010})}\BibitemShut {NoStop}%
\bibitem [{\citenamefont {Huang}\ \emph {et~al.}(2013)\citenamefont {Huang},
  \citenamefont {Gompper},\ and\ \citenamefont {Winkler}}]{huan:13}%
  \BibitemOpen
  \bibfield  {author} {\bibinfo {author} {\bibfnamefont {C.~C.}\ \bibnamefont
  {Huang}}, \bibinfo {author} {\bibfnamefont {G.}~\bibnamefont {Gompper}},\
  and\ \bibinfo {author} {\bibfnamefont {R.~G.}\ \bibnamefont {Winkler}},\
  }\bibfield  {title} {\bibinfo {title} {{E}ffect of hydrodynamic correlations
  on the dynamics of polymers in dilute solution},\ }\href
  {https://doi.org/10.1063/1.4799877} {\bibfield  {journal} {\bibinfo
  {journal} {J. Chem. Phys.}\ }\textbf {\bibinfo {volume} {138}},\ \bibinfo
  {pages} {144902} (\bibinfo {year} {2013})}\BibitemShut {NoStop}%
\bibitem [{\citenamefont {Chen}\ \emph {et~al.}(2017)\citenamefont {Chen},
  \citenamefont {Patelli}, \citenamefont {Chat{\'e}}, \citenamefont {Ma},\ and\
  \citenamefont {Shi}}]{chen:17}%
  \BibitemOpen
  \bibfield  {author} {\bibinfo {author} {\bibfnamefont {Q.-S.}\ \bibnamefont
  {Chen}}, \bibinfo {author} {\bibfnamefont {A.}~\bibnamefont {Patelli}},
  \bibinfo {author} {\bibfnamefont {H.}~\bibnamefont {Chat{\'e}}}, \bibinfo
  {author} {\bibfnamefont {Y.-Q.}\ \bibnamefont {Ma}},\ and\ \bibinfo {author}
  {\bibfnamefont {X.-Q.}\ \bibnamefont {Shi}},\ }\bibfield  {title} {\bibinfo
  {title} {Fore-aft asymmetric flocking},\ }\href
  {https://doi.org/10.1103/PhysRevE.96.020601} {\bibfield  {journal} {\bibinfo
  {journal} {Phys. Rev. E}\ }\textbf {\bibinfo {volume} {96}},\ \bibinfo
  {pages} {020601} (\bibinfo {year} {2017})}\BibitemShut {NoStop}%
\bibitem [{\citenamefont {Nikoubashman}\ and\ \citenamefont
  {Howard}(2017)}]{niko:17}%
  \BibitemOpen
  \bibfield  {author} {\bibinfo {author} {\bibfnamefont {A.}~\bibnamefont
  {Nikoubashman}}\ and\ \bibinfo {author} {\bibfnamefont {M.~P.}\ \bibnamefont
  {Howard}},\ }\bibfield  {title} {\bibinfo {title} {Equilibrium dynamics and
  shear rheology of semiflexible polymers in solution},\ }\href
  {https://doi.org/10.1021/acs.macromol.7b01876} {\bibfield  {journal}
  {\bibinfo  {journal} {Macromolecules}\ }\textbf {\bibinfo {volume} {50}},\
  \bibinfo {pages} {8279} (\bibinfo {year} {2017})}\BibitemShut {NoStop}%
\bibitem [{\citenamefont {Liebetreu}\ and\ \citenamefont
  {Likos}(2020)}]{lieb:20}%
  \BibitemOpen
  \bibfield  {author} {\bibinfo {author} {\bibfnamefont {M.}~\bibnamefont
  {Liebetreu}}\ and\ \bibinfo {author} {\bibfnamefont {C.~N.}\ \bibnamefont
  {Likos}},\ }\bibfield  {title} {\bibinfo {title} {Hydrodynamic inflation of
  ring polymers under shear},\ }\href
  {https://doi.org/10.1038/s43246-019-0006-5} {\bibfield  {journal} {\bibinfo
  {journal} {Commun. Mater.}\ }\textbf {\bibinfo {volume} {1}},\ \bibinfo
  {pages} {4} (\bibinfo {year} {2020})}\BibitemShut {NoStop}%
\bibitem [{\citenamefont {Devarajan}\ \emph {et~al.}(2022)\citenamefont
  {Devarajan}, \citenamefont {Rekhi}, \citenamefont {Nikoubashman},
  \citenamefont {Kim}, \citenamefont {Howard},\ and\ \citenamefont
  {Mittal}}]{deva:22}%
  \BibitemOpen
  \bibfield  {author} {\bibinfo {author} {\bibfnamefont {D.~S.}\ \bibnamefont
  {Devarajan}}, \bibinfo {author} {\bibfnamefont {S.}~\bibnamefont {Rekhi}},
  \bibinfo {author} {\bibfnamefont {A.}~\bibnamefont {Nikoubashman}}, \bibinfo
  {author} {\bibfnamefont {Y.~C.}\ \bibnamefont {Kim}}, \bibinfo {author}
  {\bibfnamefont {M.~P.}\ \bibnamefont {Howard}},\ and\ \bibinfo {author}
  {\bibfnamefont {J.}~\bibnamefont {Mittal}},\ }\bibfield  {title} {\bibinfo
  {title} {Effect of charge distribution on the dynamics of polyampholytic
  disordered proteins},\ }\href {https://doi.org/10.1021/acs.macromol.2c01390}
  {\bibfield  {journal} {\bibinfo  {journal} {Macromolecules}\ }\textbf
  {\bibinfo {volume} {55}},\ \bibinfo {pages} {8987} (\bibinfo {year}
  {2022})}\BibitemShut {NoStop}%
\bibitem [{\citenamefont {Wang}\ \emph
  {et~al.}(2023{\natexlab{a}})\citenamefont {Wang}, \citenamefont {Wang},
  \citenamefont {Shi}, \citenamefont {Lu},\ and\ \citenamefont {An}}]{zwan:23}%
  \BibitemOpen
  \bibfield  {author} {\bibinfo {author} {\bibfnamefont {Z.}~\bibnamefont
  {Wang}}, \bibinfo {author} {\bibfnamefont {Z.-G.}\ \bibnamefont {Wang}},
  \bibinfo {author} {\bibfnamefont {A.-C.}\ \bibnamefont {Shi}}, \bibinfo
  {author} {\bibfnamefont {Y.}~\bibnamefont {Lu}},\ and\ \bibinfo {author}
  {\bibfnamefont {L.}~\bibnamefont {An}},\ }\bibfield  {title} {\bibinfo
  {title} {Behaviors of a polymer chain in channels: From zimm to rouse
  dynamics},\ }\href {https://doi.org/10.1021/acs.macromol.3c00013} {\bibfield
  {journal} {\bibinfo  {journal} {Macromolecules}\ }\textbf {\bibinfo {volume}
  {56}},\ \bibinfo {pages} {2447} (\bibinfo {year}
  {2023}{\natexlab{a}})}\BibitemShut {NoStop}%
\bibitem [{\citenamefont {Ilg}(2022)}]{ilg:22}%
  \BibitemOpen
  \bibfield  {author} {\bibinfo {author} {\bibfnamefont {P.}~\bibnamefont
  {Ilg}},\ }\bibfield  {title} {\bibinfo {title} {Simulating the flow of
  interacting ferrofluids with multiparticle collision dynamics},\ }\href
  {https://doi.org/10.1103/PhysRevE.106.064605} {\bibfield  {journal} {\bibinfo
   {journal} {Phys. Rev. E}\ }\textbf {\bibinfo {volume} {106}},\ \bibinfo
  {pages} {064605} (\bibinfo {year} {2022})}\BibitemShut {NoStop}%
\bibitem [{\citenamefont {Wang}\ \emph
  {et~al.}(2023{\natexlab{b}})\citenamefont {Wang}, \citenamefont {Feng},
  \citenamefont {Zhang}, \citenamefont {Shao},\ and\ \citenamefont
  {Du}}]{wang:23}%
  \BibitemOpen
  \bibfield  {author} {\bibinfo {author} {\bibfnamefont {R.}~\bibnamefont
  {Wang}}, \bibinfo {author} {\bibfnamefont {C.}~\bibnamefont {Feng}}, \bibinfo
  {author} {\bibfnamefont {Z.}~\bibnamefont {Zhang}}, \bibinfo {author}
  {\bibfnamefont {C.}~\bibnamefont {Shao}},\ and\ \bibinfo {author}
  {\bibfnamefont {J.}~\bibnamefont {Du}},\ }\bibfield  {title} {\bibinfo
  {title} {What quantity of charge on the nanoparticle can result in a hybrid
  morphology of the nanofluid and a higher thermal conductivity?},\ }\href
  {https://doi.org/10.1016/j.powtec.2023.118443} {\bibfield  {journal}
  {\bibinfo  {journal} {Powder Technol.}\ }\textbf {\bibinfo {volume} {422}},\
  \bibinfo {pages} {118443} (\bibinfo {year} {2023}{\natexlab{b}})}\BibitemShut
  {NoStop}%
\bibitem [{\citenamefont {Thakur}\ and\ \citenamefont
  {Kapral}(2012)}]{thak:12}%
  \BibitemOpen
  \bibfield  {author} {\bibinfo {author} {\bibfnamefont {S.}~\bibnamefont
  {Thakur}}\ and\ \bibinfo {author} {\bibfnamefont {R.}~\bibnamefont
  {Kapral}},\ }\bibfield  {title} {\bibinfo {title} {Collective dynamics of
  self-propelled sphere-dimer motors},\ }\href
  {https://doi.org/10.1103/PhysRevE.85.026121} {\bibfield  {journal} {\bibinfo
  {journal} {Phys. Rev. E}\ }\textbf {\bibinfo {volume} {85}},\ \bibinfo
  {pages} {026121} (\bibinfo {year} {2012})}\BibitemShut {NoStop}%
\bibitem [{\citenamefont {Theers}\ \emph
  {et~al.}(2016{\natexlab{b}})\citenamefont {Theers}, \citenamefont {Westphal},
  \citenamefont {Gompper},\ and\ \citenamefont {Winkler}}]{thee:16.1}%
  \BibitemOpen
  \bibfield  {author} {\bibinfo {author} {\bibfnamefont {M.}~\bibnamefont
  {Theers}}, \bibinfo {author} {\bibfnamefont {E.}~\bibnamefont {Westphal}},
  \bibinfo {author} {\bibfnamefont {G.}~\bibnamefont {Gompper}},\ and\ \bibinfo
  {author} {\bibfnamefont {R.~G.}\ \bibnamefont {Winkler}},\ }\bibfield
  {title} {\bibinfo {title} {Modeling a spheroidal microswimmer and cooperative
  swimming in a narrow slit},\ }\href {https://doi.org/10.1039/C6SM01424K}
  {\bibfield  {journal} {\bibinfo  {journal} {Soft Matter}\ }\textbf {\bibinfo
  {volume} {12}},\ \bibinfo {pages} {7372} (\bibinfo {year}
  {2016}{\natexlab{b}})}\BibitemShut {NoStop}%
\bibitem [{\citenamefont {Zantop}\ and\ \citenamefont {Stark}(2022)}]{zant:22}%
  \BibitemOpen
  \bibfield  {author} {\bibinfo {author} {\bibfnamefont {A.~W.}\ \bibnamefont
  {Zantop}}\ and\ \bibinfo {author} {\bibfnamefont {H.}~\bibnamefont {Stark}},\
  }\bibfield  {title} {\bibinfo {title} {Emergent collective dynamics of pusher
  and puller squirmer rods: swarming, clustering, and turbulence},\ }\href
  {https://doi.org/10.1039/D2SM00449F} {\bibfield  {journal} {\bibinfo
  {journal} {Soft Matter}\ }\textbf {\bibinfo {volume} {18}},\ \bibinfo {pages}
  {6179} (\bibinfo {year} {2022})}\BibitemShut {NoStop}%
\bibitem [{\citenamefont {Qi}\ \emph {et~al.}(2022)\citenamefont {Qi},
  \citenamefont {Westphal}, \citenamefont {Gompper},\ and\ \citenamefont
  {Winkler}}]{qi:22}%
  \BibitemOpen
  \bibfield  {author} {\bibinfo {author} {\bibfnamefont {K.}~\bibnamefont
  {Qi}}, \bibinfo {author} {\bibfnamefont {E.}~\bibnamefont {Westphal}},
  \bibinfo {author} {\bibfnamefont {G.}~\bibnamefont {Gompper}},\ and\ \bibinfo
  {author} {\bibfnamefont {R.~G.}\ \bibnamefont {Winkler}},\ }\bibfield
  {title} {\bibinfo {title} {Emergence of active turbulence in microswimmer
  suspensions due to active hydrodynamic stress and volume exclusion},\ }\href
  {https://doi.org/10.1038/s42005-022-00820-7} {\bibfield  {journal} {\bibinfo
  {journal} {Commun. Phys.}\ }\textbf {\bibinfo {volume} {5}},\ \bibinfo
  {pages} {49} (\bibinfo {year} {2022})}\BibitemShut {NoStop}%
\bibitem [{\citenamefont {Goh}\ \emph {et~al.}(2023)\citenamefont {Goh},
  \citenamefont {Winkler},\ and\ \citenamefont {Gompper}}]{goh:23}%
  \BibitemOpen
  \bibfield  {author} {\bibinfo {author} {\bibfnamefont {S.}~\bibnamefont
  {Goh}}, \bibinfo {author} {\bibfnamefont {R.~G.}\ \bibnamefont {Winkler}},\
  and\ \bibinfo {author} {\bibfnamefont {G.}~\bibnamefont {Gompper}},\
  }\bibfield  {title} {\bibinfo {title} {Hydrodynamic pursuit by cognitive
  self-steering microswimmers},\ }\href
  {https://doi.org/10.1038/s42005-023-01432-5} {\bibfield  {journal} {\bibinfo
  {journal} {Commun. Phys.}\ }\textbf {\bibinfo {volume} {6}},\ \bibinfo
  {pages} {310} (\bibinfo {year} {2023})}\BibitemShut {NoStop}%
\bibitem [{\citenamefont {Macías-Durán}\ \emph {et~al.}(2023)\citenamefont
  {Macías-Durán}, \citenamefont {Duarte-Alaniz},\ and\ \citenamefont
  {Híjar}}]{dura:23}%
  \BibitemOpen
  \bibfield  {author} {\bibinfo {author} {\bibfnamefont {J.}~\bibnamefont
  {Macías-Durán}}, \bibinfo {author} {\bibfnamefont {V.}~\bibnamefont
  {Duarte-Alaniz}},\ and\ \bibinfo {author} {\bibfnamefont {H.}~\bibnamefont
  {Híjar}},\ }\bibfield  {title} {\bibinfo {title} {Active nematic liquid
  crystals simulated by particle-based mesoscopic methods},\ }\href
  {https://doi.org/10.1039/D3SM00481C} {\bibfield  {journal} {\bibinfo
  {journal} {Soft Matter}\ }\textbf {\bibinfo {volume} {19}},\ \bibinfo {pages}
  {8052} (\bibinfo {year} {2023})}\BibitemShut {NoStop}%
\bibitem [{\citenamefont {Jain}\ and\ \citenamefont {Thakur}(2022)}]{jain:22}%
  \BibitemOpen
  \bibfield  {author} {\bibinfo {author} {\bibfnamefont {N.}~\bibnamefont
  {Jain}}\ and\ \bibinfo {author} {\bibfnamefont {S.}~\bibnamefont {Thakur}},\
  }\bibfield  {title} {\bibinfo {title} {Collapse dynamics of chemically active
  flexible polymer},\ }\href {https://doi.org/10.1021/acs.macromol.1c02502}
  {\bibfield  {journal} {\bibinfo  {journal} {Macromolecules}\ }\textbf
  {\bibinfo {volume} {55}},\ \bibinfo {pages} {2375} (\bibinfo {year}
  {2022})}\BibitemShut {NoStop}%
\bibitem [{\citenamefont {Clop{\'e}s~Llah{\'\i}}\ \emph
  {et~al.}(2022)\citenamefont {Clop{\'e}s~Llah{\'\i}}, \citenamefont
  {Mart{\'\i}n-G{\'o}mez}, \citenamefont {Gompper},\ and\ \citenamefont
  {Winkler}}]{clop:22}%
  \BibitemOpen
  \bibfield  {author} {\bibinfo {author} {\bibfnamefont {J.}~\bibnamefont
  {Clop{\'e}s~Llah{\'\i}}}, \bibinfo {author} {\bibfnamefont {A.}~\bibnamefont
  {Mart{\'\i}n-G{\'o}mez}}, \bibinfo {author} {\bibfnamefont {G.}~\bibnamefont
  {Gompper}},\ and\ \bibinfo {author} {\bibfnamefont {R.~G.}\ \bibnamefont
  {Winkler}},\ }\bibfield  {title} {\bibinfo {title} {Simulating wet active
  polymers by multiparticle collision dynamics},\ }\href
  {https://doi.org/10.1103/PhysRevE.105.015310} {\bibfield  {journal} {\bibinfo
   {journal} {Phys. Rev. E}\ }\textbf {\bibinfo {volume} {105}},\ \bibinfo
  {pages} {015310} (\bibinfo {year} {2022})}\BibitemShut {NoStop}%
\bibitem [{\citenamefont {Hu}\ \emph {et~al.}(2015{\natexlab{b}})\citenamefont
  {Hu}, \citenamefont {Yang}, \citenamefont {Gompper},\ and\ \citenamefont
  {Winkler}}]{hu:15.1}%
  \BibitemOpen
  \bibfield  {author} {\bibinfo {author} {\bibfnamefont {J.}~\bibnamefont
  {Hu}}, \bibinfo {author} {\bibfnamefont {M.}~\bibnamefont {Yang}}, \bibinfo
  {author} {\bibfnamefont {G.}~\bibnamefont {Gompper}},\ and\ \bibinfo {author}
  {\bibfnamefont {R.~G.}\ \bibnamefont {Winkler}},\ }\bibfield  {title}
  {\bibinfo {title} {Modelling the mechanics and hydrodynamics of swimming {E}.
  {coli}},\ }\href {https://doi.org/10.1039/C5SM01678A} {\bibfield  {journal}
  {\bibinfo  {journal} {Soft Matter}\ }\textbf {\bibinfo {volume} {11}},\
  \bibinfo {pages} {7867} (\bibinfo {year} {2015}{\natexlab{b}})}\BibitemShut
  {NoStop}%
\bibitem [{\citenamefont {Eisenstecken}\ \emph {et~al.}(2016)\citenamefont
  {Eisenstecken}, \citenamefont {Hu},\ and\ \citenamefont
  {Winkler}}]{eise:16.1}%
  \BibitemOpen
  \bibfield  {author} {\bibinfo {author} {\bibfnamefont {T.}~\bibnamefont
  {Eisenstecken}}, \bibinfo {author} {\bibfnamefont {J.}~\bibnamefont {Hu}},\
  and\ \bibinfo {author} {\bibfnamefont {R.~G.}\ \bibnamefont {Winkler}},\
  }\bibfield  {title} {\bibinfo {title} {Bacterial swarmer cells in
  confinement: a mesoscale hydrodynamic simulation study},\ }\href
  {https://doi.org/10.1039/C6SM01532H} {\bibfield  {journal} {\bibinfo
  {journal} {Soft Matter}\ }\textbf {\bibinfo {volume} {12}},\ \bibinfo {pages}
  {8316} (\bibinfo {year} {2016})}\BibitemShut {NoStop}%
\bibitem [{\citenamefont {Ning}\ \emph {et~al.}(2023)\citenamefont {Ning},
  \citenamefont {Lou}, \citenamefont {Ma}, \citenamefont {Yang}, \citenamefont
  {Luo}, \citenamefont {Chen}, \citenamefont {Meng}, \citenamefont {Zhou},
  \citenamefont {Yang},\ and\ \citenamefont {Peng}}]{ning:23}%
  \BibitemOpen
  \bibfield  {author} {\bibinfo {author} {\bibfnamefont {L.}~\bibnamefont
  {Ning}}, \bibinfo {author} {\bibfnamefont {X.}~\bibnamefont {Lou}}, \bibinfo
  {author} {\bibfnamefont {Q.}~\bibnamefont {Ma}}, \bibinfo {author}
  {\bibfnamefont {Y.}~\bibnamefont {Yang}}, \bibinfo {author} {\bibfnamefont
  {N.}~\bibnamefont {Luo}}, \bibinfo {author} {\bibfnamefont {K.}~\bibnamefont
  {Chen}}, \bibinfo {author} {\bibfnamefont {F.}~\bibnamefont {Meng}}, \bibinfo
  {author} {\bibfnamefont {X.}~\bibnamefont {Zhou}}, \bibinfo {author}
  {\bibfnamefont {M.}~\bibnamefont {Yang}},\ and\ \bibinfo {author}
  {\bibfnamefont {Y.}~\bibnamefont {Peng}},\ }\bibfield  {title} {\bibinfo
  {title} {Hydrodynamics-induced long-range attraction between plates in
  bacterial suspensions},\ }\href
  {https://doi.org/10.1103/PhysRevLett.131.158301} {\bibfield  {journal}
  {\bibinfo  {journal} {Phys. Rev. Lett.}\ }\textbf {\bibinfo {volume} {131}},\
  \bibinfo {pages} {158301} (\bibinfo {year} {2023})}\BibitemShut {NoStop}%
\bibitem [{\citenamefont {Elgeti}\ \emph {et~al.}(2010)\citenamefont {Elgeti},
  \citenamefont {Kaupp},\ and\ \citenamefont {Gompper}}]{elge:10}%
  \BibitemOpen
  \bibfield  {author} {\bibinfo {author} {\bibfnamefont {J.}~\bibnamefont
  {Elgeti}}, \bibinfo {author} {\bibfnamefont {U.~B.}\ \bibnamefont {Kaupp}},\
  and\ \bibinfo {author} {\bibfnamefont {G.}~\bibnamefont {Gompper}},\
  }\bibfield  {title} {\bibinfo {title} {Hydrodynamics of sperm cells near
  surfaces},\ }\href {https://doi.org/10.1016/j.bpj.2010.05.015} {\bibfield
  {journal} {\bibinfo  {journal} {Biophys. J.}\ }\textbf {\bibinfo {volume}
  {99}},\ \bibinfo {pages} {1018} (\bibinfo {year} {2010})}\BibitemShut
  {NoStop}%
\bibitem [{\citenamefont {Chinnasamy}\ \emph {et~al.}(2018)\citenamefont
  {Chinnasamy}, \citenamefont {Kingsley}, \citenamefont {Inci}, \citenamefont
  {Turek}, \citenamefont {Rosen}, \citenamefont {Behr}, \citenamefont
  {Tüzel},\ and\ \citenamefont {Demirci}}]{chin:18}%
  \BibitemOpen
  \bibfield  {author} {\bibinfo {author} {\bibfnamefont {T.}~\bibnamefont
  {Chinnasamy}}, \bibinfo {author} {\bibfnamefont {J.~L.}\ \bibnamefont
  {Kingsley}}, \bibinfo {author} {\bibfnamefont {F.}~\bibnamefont {Inci}},
  \bibinfo {author} {\bibfnamefont {P.~J.}\ \bibnamefont {Turek}}, \bibinfo
  {author} {\bibfnamefont {M.~P.}\ \bibnamefont {Rosen}}, \bibinfo {author}
  {\bibfnamefont {B.}~\bibnamefont {Behr}}, \bibinfo {author} {\bibfnamefont
  {E.}~\bibnamefont {Tüzel}},\ and\ \bibinfo {author} {\bibfnamefont
  {U.}~\bibnamefont {Demirci}},\ }\bibfield  {title} {\bibinfo {title}
  {Guidance and self-sorting of active swimmers: 3d periodic arrays increase
  persistence length of human sperm selecting for the fittest},\ }\href
  {https://doi.org/10.1002/advs.201700531} {\bibfield  {journal} {\bibinfo
  {journal} {Adv. Sci.}\ }\textbf {\bibinfo {volume} {5}},\ \bibinfo {pages}
  {1700531} (\bibinfo {year} {2018})}\BibitemShut {NoStop}%
\bibitem [{\citenamefont {Rode}\ \emph {et~al.}(2019)\citenamefont {Rode},
  \citenamefont {Elgeti},\ and\ \citenamefont {Gompper}}]{rode:19}%
  \BibitemOpen
  \bibfield  {author} {\bibinfo {author} {\bibfnamefont {S.}~\bibnamefont
  {Rode}}, \bibinfo {author} {\bibfnamefont {J.}~\bibnamefont {Elgeti}},\ and\
  \bibinfo {author} {\bibfnamefont {G.}~\bibnamefont {Gompper}},\ }\bibfield
  {title} {\bibinfo {title} {Sperm motility in modulated microchannels},\
  }\href {https://doi.org/10.1088/1367-2630/aaf544} {\bibfield  {journal}
  {\bibinfo  {journal} {New J. Phys.}\ }\textbf {\bibinfo {volume} {21}},\
  \bibinfo {pages} {013016} (\bibinfo {year} {2019})}\BibitemShut {NoStop}%
\bibitem [{\citenamefont {Heddergott}\ \emph {et~al.}(2012)\citenamefont
  {Heddergott}, \citenamefont {Krüger}, \citenamefont {Babu}, \citenamefont
  {Wei}, \citenamefont {Stellamanns}, \citenamefont {Uppaluri}, \citenamefont
  {Pfohl}, \citenamefont {Stark},\ and\ \citenamefont {Engstler}}]{hedd:12}%
  \BibitemOpen
  \bibfield  {author} {\bibinfo {author} {\bibfnamefont {N.}~\bibnamefont
  {Heddergott}}, \bibinfo {author} {\bibfnamefont {T.}~\bibnamefont {Krüger}},
  \bibinfo {author} {\bibfnamefont {S.~B.}\ \bibnamefont {Babu}}, \bibinfo
  {author} {\bibfnamefont {A.}~\bibnamefont {Wei}}, \bibinfo {author}
  {\bibfnamefont {E.}~\bibnamefont {Stellamanns}}, \bibinfo {author}
  {\bibfnamefont {S.}~\bibnamefont {Uppaluri}}, \bibinfo {author}
  {\bibfnamefont {T.}~\bibnamefont {Pfohl}}, \bibinfo {author} {\bibfnamefont
  {H.}~\bibnamefont {Stark}},\ and\ \bibinfo {author} {\bibfnamefont
  {M.}~\bibnamefont {Engstler}},\ }\bibfield  {title} {\bibinfo {title}
  {Trypanosome motion represents an adaptation to the crowded environment of
  the vertebrate bloodstream},\ }\href
  {https://doi.org/10.1371/journal.ppat.1003023} {\bibfield  {journal}
  {\bibinfo  {journal} {PLOS Pathog.}\ }\textbf {\bibinfo {volume} {8}},\
  \bibinfo {pages} {1} (\bibinfo {year} {2012})}\BibitemShut {NoStop}%
\bibitem [{\citenamefont {Lansche}\ \emph {et~al.}(2018)\citenamefont
  {Lansche}, \citenamefont {Dasanna}, \citenamefont {Quadt}, \citenamefont
  {Fröhlich}, \citenamefont {Missirlis}, \citenamefont {Tétard},
  \citenamefont {Gamain}, \citenamefont {Buchholz}, \citenamefont {Sanchez},
  \citenamefont {Tanaka}, \citenamefont {Schwarz},\ and\ \citenamefont
  {Lanzer}}]{lans:18}%
  \BibitemOpen
  \bibfield  {author} {\bibinfo {author} {\bibfnamefont {C.}~\bibnamefont
  {Lansche}}, \bibinfo {author} {\bibfnamefont {A.~K.}\ \bibnamefont
  {Dasanna}}, \bibinfo {author} {\bibfnamefont {K.}~\bibnamefont {Quadt}},
  \bibinfo {author} {\bibfnamefont {B.}~\bibnamefont {Fröhlich}}, \bibinfo
  {author} {\bibfnamefont {D.}~\bibnamefont {Missirlis}}, \bibinfo {author}
  {\bibfnamefont {M.}~\bibnamefont {Tétard}}, \bibinfo {author} {\bibfnamefont
  {B.}~\bibnamefont {Gamain}}, \bibinfo {author} {\bibfnamefont
  {B.}~\bibnamefont {Buchholz}}, \bibinfo {author} {\bibfnamefont {C.~P.}\
  \bibnamefont {Sanchez}}, \bibinfo {author} {\bibfnamefont {M.}~\bibnamefont
  {Tanaka}}, \bibinfo {author} {\bibfnamefont {U.~S.}\ \bibnamefont
  {Schwarz}},\ and\ \bibinfo {author} {\bibfnamefont {M.}~\bibnamefont
  {Lanzer}},\ }\bibfield  {title} {\bibinfo {title} {The sickle cell trait
  affects contact dynamics and endothelial cell activation in \emph{Plasmodium
  falciparum}-infected erythrocytes},\ }\href
  {https://doi.org/10.1038/s42003-018-0223-3} {\bibfield  {journal} {\bibinfo
  {journal} {Commun. Biol.}\ }\textbf {\bibinfo {volume} {1}},\ \bibinfo
  {pages} {211} (\bibinfo {year} {2018})}\BibitemShut {NoStop}%
\bibitem [{\citenamefont {Ihle}\ and\ \citenamefont {Kroll}(2001)}]{ihle:01}%
  \BibitemOpen
  \bibfield  {author} {\bibinfo {author} {\bibfnamefont {T.}~\bibnamefont
  {Ihle}}\ and\ \bibinfo {author} {\bibfnamefont {D.~M.}\ \bibnamefont
  {Kroll}},\ }\bibfield  {title} {\bibinfo {title} {Stochastic rotation
  dynamics: A {Galilean}-invariant mesoscopic model for fluid flow},\ }\href
  {https://doi.org/10.1103/PhysRevE.63.020201} {\bibfield  {journal} {\bibinfo
  {journal} {Phys. Rev. E}\ }\textbf {\bibinfo {volume} {63}},\ \bibinfo
  {pages} {020201(R)} (\bibinfo {year} {2001})}\BibitemShut {NoStop}%
\bibitem [{\citenamefont {G{\"o}tze}\ \emph {et~al.}(2007)\citenamefont
  {G{\"o}tze}, \citenamefont {Noguchi},\ and\ \citenamefont
  {Gompper}}]{goet:07}%
  \BibitemOpen
  \bibfield  {author} {\bibinfo {author} {\bibfnamefont {I.~O.}\ \bibnamefont
  {G{\"o}tze}}, \bibinfo {author} {\bibfnamefont {H.}~\bibnamefont {Noguchi}},\
  and\ \bibinfo {author} {\bibfnamefont {G.}~\bibnamefont {Gompper}},\
  }\bibfield  {title} {\bibinfo {title} {Relevance of angular momentum
  conservation in mesoscale hydrodynamics simulations},\ }\href
  {https://doi.org/10.1103/PhysRevE.76.046705} {\bibfield  {journal} {\bibinfo
  {journal} {Phys. Rev. E}\ }\textbf {\bibinfo {volume} {76}},\ \bibinfo
  {pages} {046705} (\bibinfo {year} {2007})}\BibitemShut {NoStop}%
\bibitem [{\citenamefont {Theers}\ and\ \citenamefont
  {Winkler}(2015)}]{thee:15}%
  \BibitemOpen
  \bibfield  {author} {\bibinfo {author} {\bibfnamefont {M.}~\bibnamefont
  {Theers}}\ and\ \bibinfo {author} {\bibfnamefont {R.~G.}\ \bibnamefont
  {Winkler}},\ }\bibfield  {title} {\bibinfo {title} {Bulk viscosity of
  multiparticle collision dynamics fluids},\ }\href
  {https://doi.org/10.1103/PhysRevE.91.033309} {\bibfield  {journal} {\bibinfo
  {journal} {Phys. Rev. E}\ }\textbf {\bibinfo {volume} {91}},\ \bibinfo
  {pages} {033309} (\bibinfo {year} {2015})}\BibitemShut {NoStop}%
\bibitem [{\citenamefont {Yang}\ \emph {et~al.}(2015)\citenamefont {Yang},
  \citenamefont {Theers}, \citenamefont {Hu}, \citenamefont {Gompper},
  \citenamefont {Winkler},\ and\ \citenamefont {Ripoll}}]{yang:15}%
  \BibitemOpen
  \bibfield  {author} {\bibinfo {author} {\bibfnamefont {M.}~\bibnamefont
  {Yang}}, \bibinfo {author} {\bibfnamefont {M.}~\bibnamefont {Theers}},
  \bibinfo {author} {\bibfnamefont {J.}~\bibnamefont {Hu}}, \bibinfo {author}
  {\bibfnamefont {G.}~\bibnamefont {Gompper}}, \bibinfo {author} {\bibfnamefont
  {R.~G.}\ \bibnamefont {Winkler}},\ and\ \bibinfo {author} {\bibfnamefont
  {M.}~\bibnamefont {Ripoll}},\ }\bibfield  {title} {\bibinfo {title} {Effect
  of angular momentum conservation on hydrodynamic simulations of colloids},\
  }\href {https://doi.org/10.1103/PhysRevE.92.013301} {\bibfield  {journal}
  {\bibinfo  {journal} {Phys. Rev. E}\ }\textbf {\bibinfo {volume} {92}},\
  \bibinfo {pages} {013301} (\bibinfo {year} {2015})}\BibitemShut {NoStop}%
\bibitem [{\citenamefont {Noguchi}\ \emph {et~al.}(2007)\citenamefont
  {Noguchi}, \citenamefont {Kikuchi},\ and\ \citenamefont {Gompper}}]{nogu:07}%
  \BibitemOpen
  \bibfield  {author} {\bibinfo {author} {\bibfnamefont {H.}~\bibnamefont
  {Noguchi}}, \bibinfo {author} {\bibfnamefont {N.}~\bibnamefont {Kikuchi}},\
  and\ \bibinfo {author} {\bibfnamefont {G.}~\bibnamefont {Gompper}},\
  }\bibfield  {title} {\bibinfo {title} {Particle-based mesoscale hydrodynamic
  techniques},\ }\href {https://doi.org/10.1209/0295-5075/78/10005} {\bibfield
  {journal} {\bibinfo  {journal} {EPL}\ }\textbf {\bibinfo {volume} {78}},\
  \bibinfo {pages} {10005} (\bibinfo {year} {2007})}\BibitemShut {NoStop}%
\bibitem [{\citenamefont {Huang}\ \emph
  {et~al.}(2010{\natexlab{b}})\citenamefont {Huang}, \citenamefont {Chatterji},
  \citenamefont {Sutmann}, \citenamefont {Gompper},\ and\ \citenamefont
  {Winkler}}]{huan:10.1}%
  \BibitemOpen
  \bibfield  {author} {\bibinfo {author} {\bibfnamefont {C.-C.}\ \bibnamefont
  {Huang}}, \bibinfo {author} {\bibfnamefont {A.}~\bibnamefont {Chatterji}},
  \bibinfo {author} {\bibfnamefont {G.}~\bibnamefont {Sutmann}}, \bibinfo
  {author} {\bibfnamefont {G.}~\bibnamefont {Gompper}},\ and\ \bibinfo {author}
  {\bibfnamefont {R.~G.}\ \bibnamefont {Winkler}},\ }\bibfield  {title}
  {\bibinfo {title} {Cell-level canonical sampling by velocity scaling for
  multiparticle collision dynamics simulations},\ }\href
  {https://doi.org/10.1016/j.jcp.2009.09.024} {\bibfield  {journal} {\bibinfo
  {journal} {J. Comput. Phys.}\ }\textbf {\bibinfo {volume} {229}},\ \bibinfo
  {pages} {168} (\bibinfo {year} {2010}{\natexlab{b}})}\BibitemShut {NoStop}%
\bibitem [{\citenamefont {Lamura}\ \emph {et~al.}(2001)\citenamefont {Lamura},
  \citenamefont {Gompper}, \citenamefont {Ihle},\ and\ \citenamefont
  {Kroll}}]{lamu:01}%
  \BibitemOpen
  \bibfield  {author} {\bibinfo {author} {\bibfnamefont {A.}~\bibnamefont
  {Lamura}}, \bibinfo {author} {\bibfnamefont {G.}~\bibnamefont {Gompper}},
  \bibinfo {author} {\bibfnamefont {T.}~\bibnamefont {Ihle}},\ and\ \bibinfo
  {author} {\bibfnamefont {D.~M.}\ \bibnamefont {Kroll}},\ }\bibfield  {title}
  {\bibinfo {title} {Multiparticle collision dynamics: Flow around a circular
  and a square cylinder},\ }\href {https://doi.org/10.1209/epl/i2001-00522-9}
  {\bibfield  {journal} {\bibinfo  {journal} {Europhys. Lett.}\ }\textbf
  {\bibinfo {volume} {56}},\ \bibinfo {pages} {319} (\bibinfo {year}
  {2001})}\BibitemShut {NoStop}%
\bibitem [{\citenamefont {Cannavacciuolo}\ \emph {et~al.}(2008)\citenamefont
  {Cannavacciuolo}, \citenamefont {Winkler},\ and\ \citenamefont
  {Gompper}}]{cann:08}%
  \BibitemOpen
  \bibfield  {author} {\bibinfo {author} {\bibfnamefont {L.}~\bibnamefont
  {Cannavacciuolo}}, \bibinfo {author} {\bibfnamefont {R.~G.}\ \bibnamefont
  {Winkler}},\ and\ \bibinfo {author} {\bibfnamefont {G.}~\bibnamefont
  {Gompper}},\ }\bibfield  {title} {\bibinfo {title} {Mesoscale simulation of
  polymer dynamics in microchannel flows},\ }\href
  {https://doi.org/10.1209/0295-5075/83/34007} {\bibfield  {journal} {\bibinfo
  {journal} {EPL}\ }\textbf {\bibinfo {volume} {83}},\ \bibinfo {pages} {34007}
  (\bibinfo {year} {2008})}\BibitemShut {NoStop}%
\bibitem [{\citenamefont {Huang}\ \emph {et~al.}(2015)\citenamefont {Huang},
  \citenamefont {Varghese}, \citenamefont {Gompper},\ and\ \citenamefont
  {Winkler}}]{huan:15}%
  \BibitemOpen
  \bibfield  {author} {\bibinfo {author} {\bibfnamefont {C.-C.}\ \bibnamefont
  {Huang}}, \bibinfo {author} {\bibfnamefont {A.}~\bibnamefont {Varghese}},
  \bibinfo {author} {\bibfnamefont {G.}~\bibnamefont {Gompper}},\ and\ \bibinfo
  {author} {\bibfnamefont {R.~G.}\ \bibnamefont {Winkler}},\ }\bibfield
  {title} {\bibinfo {title} {Thermostat for nonequilibrium
  multiparticle-collision-dynamics simulations},\ }\href
  {https://doi.org/10.1103/PhysRevE.91.013310} {\bibfield  {journal} {\bibinfo
  {journal} {Phys. Rev. E}\ }\textbf {\bibinfo {volume} {91}},\ \bibinfo
  {pages} {013310} (\bibinfo {year} {2015})}\BibitemShut {NoStop}%
\bibitem [{\citenamefont {Ripoll}\ \emph {et~al.}(2005)\citenamefont {Ripoll},
  \citenamefont {Mussawisade}, \citenamefont {Winkler},\ and\ \citenamefont
  {Gompper}}]{ripo:05}%
  \BibitemOpen
  \bibfield  {author} {\bibinfo {author} {\bibfnamefont {M.}~\bibnamefont
  {Ripoll}}, \bibinfo {author} {\bibfnamefont {K.}~\bibnamefont {Mussawisade}},
  \bibinfo {author} {\bibfnamefont {R.~G.}\ \bibnamefont {Winkler}},\ and\
  \bibinfo {author} {\bibfnamefont {G.}~\bibnamefont {Gompper}},\ }\bibfield
  {title} {\bibinfo {title} {Dynamic regimes of fluids simulated by
  multi-particle-collision dynamics},\ }\href
  {https://doi.org/10.1103/PhysRevE.72.016701} {\bibfield  {journal} {\bibinfo
  {journal} {Phys. Rev. E}\ }\textbf {\bibinfo {volume} {72}},\ \bibinfo
  {pages} {016701} (\bibinfo {year} {2005})}\BibitemShut {NoStop}%
\bibitem [{\citenamefont {Theers}\ \emph {et~al.}(2018)\citenamefont {Theers},
  \citenamefont {Westphal}, \citenamefont {Qi}, \citenamefont {Winkler},\ and\
  \citenamefont {Gompper}}]{thee:18}%
  \BibitemOpen
  \bibfield  {author} {\bibinfo {author} {\bibfnamefont {M.}~\bibnamefont
  {Theers}}, \bibinfo {author} {\bibfnamefont {E.}~\bibnamefont {Westphal}},
  \bibinfo {author} {\bibfnamefont {K.}~\bibnamefont {Qi}}, \bibinfo {author}
  {\bibfnamefont {R.~G.}\ \bibnamefont {Winkler}},\ and\ \bibinfo {author}
  {\bibfnamefont {G.}~\bibnamefont {Gompper}},\ }\bibfield  {title} {\bibinfo
  {title} {Clustering of microswimmers: interplay of shape and hydrodynamics},\
  }\href {https://doi.org/10.1039/C8SM01390J} {\bibfield  {journal} {\bibinfo
  {journal} {Soft Matter}\ }\textbf {\bibinfo {volume} {14}},\ \bibinfo {pages}
  {8590} (\bibinfo {year} {2018})}\BibitemShut {NoStop}%
\bibitem [{\citenamefont {Goldberg}(1991)}]{gold:91}%
  \BibitemOpen
  \bibfield  {author} {\bibinfo {author} {\bibfnamefont {D.}~\bibnamefont
  {Goldberg}},\ }\bibfield  {title} {\bibinfo {title} {What every computer
  scientist should know about floating-point arithmetic},\ }\href
  {https://doi.org/10.1145/103162.103163} {\bibfield  {journal} {\bibinfo
  {journal} {ACM comput. surv. (CSUR)}\ }\textbf {\bibinfo {volume} {23}},\
  \bibinfo {pages} {5} (\bibinfo {year} {1991})}\BibitemShut {NoStop}%
\bibitem [{\citenamefont {Villa}\ \emph {et~al.}(2009)\citenamefont {Villa},
  \citenamefont {Chavarria-Miranda}, \citenamefont {Gurumoorthi}, \citenamefont
  {M{\'a}rquez},\ and\ \citenamefont {Krishnamoorthy}}]{vill:09}%
  \BibitemOpen
  \bibfield  {author} {\bibinfo {author} {\bibfnamefont {O.}~\bibnamefont
  {Villa}}, \bibinfo {author} {\bibfnamefont {D.}~\bibnamefont
  {Chavarria-Miranda}}, \bibinfo {author} {\bibfnamefont {V.}~\bibnamefont
  {Gurumoorthi}}, \bibinfo {author} {\bibfnamefont {A.}~\bibnamefont
  {M{\'a}rquez}},\ and\ \bibinfo {author} {\bibfnamefont {S.}~\bibnamefont
  {Krishnamoorthy}},\ }\bibfield  {title} {\bibinfo {title} {Effects of
  floating-point non-associativity on numerical computations on massively
  multithreaded systems},\ }in\ \href@noop {} {\emph {\bibinfo {booktitle}
  {Proceedings of Cray User Group Meeting (CUG)}}},\ Vol.~\bibinfo {volume}
  {3}\ (\bibinfo {year} {2009})\BibitemShut {NoStop}%
\bibitem [{\citenamefont {{Le Grand}}\ \emph {et~al.}(2013)\citenamefont {{Le
  Grand}}, \citenamefont {Götz},\ and\ \citenamefont {Walker}}]{gran:13}%
  \BibitemOpen
  \bibfield  {author} {\bibinfo {author} {\bibfnamefont {S.}~\bibnamefont {{Le
  Grand}}}, \bibinfo {author} {\bibfnamefont {A.~W.}\ \bibnamefont {Götz}},\
  and\ \bibinfo {author} {\bibfnamefont {R.~C.}\ \bibnamefont {Walker}},\
  }\bibfield  {title} {\bibinfo {title} {Spfp: Speed without compromise—a
  mixed precision model for gpu accelerated molecular dynamics simulations},\
  }\href {https://doi.org/10.1016/j.cpc.2012.09.022} {\bibfield  {journal}
  {\bibinfo  {journal} {Comput. Phys. Commun.}\ }\textbf {\bibinfo {volume}
  {184}},\ \bibinfo {pages} {374} (\bibinfo {year} {2013})}\BibitemShut
  {NoStop}%
\bibitem [{\citenamefont {Dietz}\ \emph {et~al.}(2015)\citenamefont {Dietz},
  \citenamefont {Li}, \citenamefont {Regehr},\ and\ \citenamefont
  {Adve}}]{diet:15}%
  \BibitemOpen
  \bibfield  {author} {\bibinfo {author} {\bibfnamefont {W.}~\bibnamefont
  {Dietz}}, \bibinfo {author} {\bibfnamefont {P.}~\bibnamefont {Li}}, \bibinfo
  {author} {\bibfnamefont {J.}~\bibnamefont {Regehr}},\ and\ \bibinfo {author}
  {\bibfnamefont {V.}~\bibnamefont {Adve}},\ }\bibfield  {title} {\bibinfo
  {title} {Understanding integer overflow in c/c++},\ }\bibfield  {journal}
  {\bibinfo  {journal} {ACM Trans. Softw. Eng. Methodol.}\ }\textbf {\bibinfo
  {volume} {25}},\ \href {https://doi.org/10.1145/2743019} {10.1145/2743019}
  (\bibinfo {year} {2015})\BibitemShut {NoStop}%
\bibitem [{\citenamefont {Westphal}(2015)}]{west:15}%
  \BibitemOpen
  \bibfield  {author} {\bibinfo {author} {\bibfnamefont {E.}~\bibnamefont
  {Westphal}},\ }\href@noop {} {\bibinfo {title} {Voting and shuffling to
  optimize atomic operations}} (\bibinfo {year} {2015}),\ \bibinfo {note}
  {{T}echnical Blog (GPU Pro Tip), NVIDIA developer,
  \url{https://developer.nvidia.com/blog/voting-and-shuffling-optimize-atomic-operations}}\BibitemShut
  {NoStop}%
\bibitem [{\citenamefont {Marsaglia}(2003)}]{mars:03}%
  \BibitemOpen
  \bibfield  {author} {\bibinfo {author} {\bibfnamefont {G.}~\bibnamefont
  {Marsaglia}},\ }\bibfield  {title} {\bibinfo {title} {Random number
  generators},\ }\href {https://doi.org/10.22237/jmasm/1051747320} {\bibfield
  {journal} {\bibinfo  {journal} {JMASM}\ }\textbf {\bibinfo {volume} {2}},\
  \bibinfo {pages} {2} (\bibinfo {year} {2003})}\BibitemShut {NoStop}%
\bibitem [{\citenamefont {{J\"{u}lich Supercomputing Centre}}(2021)}]{juwe:21}%
  \BibitemOpen
  \bibfield  {author} {\bibinfo {author} {\bibnamefont {{J\"{u}lich
  Supercomputing Centre}}},\ }\bibfield  {title} {\bibinfo {title} {{JUWELS
  Cluster and Booster: Exascale Pathfinder with Modular Supercomputing
  Architecture at Juelich Supercomputing Centre}},\ }\href
  {https://doi.org/10.17815/jlsrf-7-183} {\bibfield  {journal} {\bibinfo
  {journal} {JLSRF}\ }\textbf {\bibinfo {volume} {7}},\ \bibinfo {pages} {A138}
  (\bibinfo {year} {2021})}\BibitemShut {NoStop}%
\bibitem [{\citenamefont {Lighthill}(1952)}]{ligh:52}%
  \BibitemOpen
  \bibfield  {author} {\bibinfo {author} {\bibfnamefont {M.~J.}\ \bibnamefont
  {Lighthill}},\ }\bibfield  {title} {\bibinfo {title} {On the squirming motion
  of nearly spherical deformable bodies through liquids at very small
  {R}eynolds numbers},\ }\href {https://doi.org/10.1002/cpa.3160050201}
  {\bibfield  {journal} {\bibinfo  {journal} {Comm. Pure Appl. Math.}\ }\textbf
  {\bibinfo {volume} {5}},\ \bibinfo {pages} {109} (\bibinfo {year}
  {1952})}\BibitemShut {NoStop}%
\bibitem [{\citenamefont {Blake}(1971)}]{blak:71}%
  \BibitemOpen
  \bibfield  {author} {\bibinfo {author} {\bibfnamefont {J.~R.}\ \bibnamefont
  {Blake}},\ }\bibfield  {title} {\bibinfo {title} {A spherical envelope
  approach to ciliary propulsion},\ }\href
  {https://doi.org/10.1017/S002211207100048X} {\bibfield  {journal} {\bibinfo
  {journal} {J. Fluid Mech.}\ }\textbf {\bibinfo {volume} {46}},\ \bibinfo
  {pages} {199} (\bibinfo {year} {1971})}\BibitemShut {NoStop}%
\bibitem [{\citenamefont {G{\"o}tze}\ and\ \citenamefont
  {Gompper}(2010)}]{goet:10}%
  \BibitemOpen
  \bibfield  {author} {\bibinfo {author} {\bibfnamefont {I.~O.}\ \bibnamefont
  {G{\"o}tze}}\ and\ \bibinfo {author} {\bibfnamefont {G.}~\bibnamefont
  {Gompper}},\ }\bibfield  {title} {\bibinfo {title} {Mesoscale simulations of
  hydrodynamic squirmer interactions},\ }\href
  {https://doi.org/10.1103/PhysRevE.82.041921} {\bibfield  {journal} {\bibinfo
  {journal} {Phys. Rev. E}\ }\textbf {\bibinfo {volume} {82}},\ \bibinfo
  {pages} {041921} (\bibinfo {year} {2010})}\BibitemShut {NoStop}%
\bibitem [{\citenamefont {Pak}\ and\ \citenamefont {Lauga}(2014)}]{pak:14}%
  \BibitemOpen
  \bibfield  {author} {\bibinfo {author} {\bibfnamefont {O.~S.}\ \bibnamefont
  {Pak}}\ and\ \bibinfo {author} {\bibfnamefont {E.}~\bibnamefont {Lauga}},\
  }\bibfield  {title} {\bibinfo {title} {Generalized squirming motion of a
  sphere},\ }\href {https://doi.org/10.1007/s10665-014-9690-9} {\bibfield
  {journal} {\bibinfo  {journal} {J. Eng. Math.}\ }\textbf {\bibinfo {volume}
  {88}},\ \bibinfo {pages} {1} (\bibinfo {year} {2014})}\BibitemShut {NoStop}%
\bibitem [{\citenamefont {Wysocki}\ \emph {et~al.}(2014)\citenamefont
  {Wysocki}, \citenamefont {Winkler},\ and\ \citenamefont {Gompper}}]{wyso:14}%
  \BibitemOpen
  \bibfield  {author} {\bibinfo {author} {\bibfnamefont {A.}~\bibnamefont
  {Wysocki}}, \bibinfo {author} {\bibfnamefont {R.~G.}\ \bibnamefont
  {Winkler}},\ and\ \bibinfo {author} {\bibfnamefont {G.}~\bibnamefont
  {Gompper}},\ }\bibfield  {title} {\bibinfo {title} {Cooperative motion of
  active {B}rownian spheres in three-dimensional dense suspensions},\ }\href
  {https://doi.org/10.1209/0295-5075/105/48004} {\bibfield  {journal} {\bibinfo
   {journal} {EPL}\ }\textbf {\bibinfo {volume} {105}},\ \bibinfo {pages}
  {48004} (\bibinfo {year} {2014})}\BibitemShut {NoStop}%
\bibitem [{\citenamefont {Stenhammar}\ \emph {et~al.}(2014)\citenamefont
  {Stenhammar}, \citenamefont {Marenduzzo}, \citenamefont {Allen},\ and\
  \citenamefont {Cates}}]{sten:14}%
  \BibitemOpen
  \bibfield  {author} {\bibinfo {author} {\bibfnamefont {J.}~\bibnamefont
  {Stenhammar}}, \bibinfo {author} {\bibfnamefont {D.}~\bibnamefont
  {Marenduzzo}}, \bibinfo {author} {\bibfnamefont {R.~J.}\ \bibnamefont
  {Allen}},\ and\ \bibinfo {author} {\bibfnamefont {M.~E.}\ \bibnamefont
  {Cates}},\ }\bibfield  {title} {\bibinfo {title} {Phase behaviour of active
  {Brownian} particles: the role of dimensionality},\ }\href
  {https://doi.org/10.1039/C3SM52813H} {\bibfield  {journal} {\bibinfo
  {journal} {Soft Matter}\ }\textbf {\bibinfo {volume} {10}},\ \bibinfo {pages}
  {1489} (\bibinfo {year} {2014})}\BibitemShut {NoStop}%
\bibitem [{\citenamefont {Rycroft}(2009)}]{rycr:09}%
  \BibitemOpen
  \bibfield  {author} {\bibinfo {author} {\bibfnamefont {C.~H.}\ \bibnamefont
  {Rycroft}},\ }\bibfield  {title} {\bibinfo {title} {{VORO}++: {A}
  three-dimensional {Voronoi} cell library in {C}++},\ }\href
  {https://doi.org/10.1063/1.3215722} {\bibfield  {journal} {\bibinfo
  {journal} {Chaos}\ }\textbf {\bibinfo {volume} {19}},\ \bibinfo {pages}
  {041111} (\bibinfo {year} {2009})}\BibitemShut {NoStop}%
\bibitem [{\citenamefont {Omar}\ \emph {et~al.}(2021)\citenamefont {Omar},
  \citenamefont {Klymko}, \citenamefont {GrandPre},\ and\ \citenamefont
  {Geissler}}]{omar:21}%
  \BibitemOpen
  \bibfield  {author} {\bibinfo {author} {\bibfnamefont {A.~K.}\ \bibnamefont
  {Omar}}, \bibinfo {author} {\bibfnamefont {K.}~\bibnamefont {Klymko}},
  \bibinfo {author} {\bibfnamefont {T.}~\bibnamefont {GrandPre}},\ and\
  \bibinfo {author} {\bibfnamefont {P.~L.}\ \bibnamefont {Geissler}},\
  }\bibfield  {title} {\bibinfo {title} {Phase diagram of active brownian
  spheres: Crystallization and the metastability of motility-induced phase
  separation},\ }\href {https://doi.org/10.1103/PhysRevLett.126.188002}
  {\bibfield  {journal} {\bibinfo  {journal} {Phys. Rev. Lett.}\ }\textbf
  {\bibinfo {volume} {126}},\ \bibinfo {pages} {188002} (\bibinfo {year}
  {2021})}\BibitemShut {NoStop}%
\end{thebibliography}

\end{document}